\documentclass[]{aa} 
\usepackage[fleqn]{mathtools} 
\usepackage[varg]{txfonts}
\usepackage{natbib,twoopt}
\usepackage[colorlinks=true,urlcolor=blue,citecolor=blue,linkcolor=blue]{hyperref}
\bibpunct{(}{)}{;}{a}{}{,} 

\makeatletter
 \newcommandtwoopt{\citeads}[3][][]{%
   \nonstopmode
   \href{http://adsabs.harvard.edu/abs/#3}%
        {\def\hyper@linkstart##1##2{}%
         \let\hyper@linkend\@empty\citealp[#1][#2]{#3}}
   \biblink{#3}{\href{http://adsabs.harvard.edu/abs/#3}{ADS}}%
   \errorstopmode}            
 \newcommandtwoopt{\citepads}[3][][]{%
   \nonstopmode
   \href{http://adsabs.harvard.edu/abs/#3}%
        {\def\hyper@linkstart##1##2{}%
         \let\hyper@linkend\@empty\citep[#1][#2]{#3}}
   \biblink{#3}{\href{http://adsabs.harvard.edu/abs/#3}{ADS}}
   \errorstopmode}            
 \newcommandtwoopt{\citetads}[3][][]{%
   \nonstopmode
   \href{http://adsabs.harvard.edu/abs/#3}
        {\def\hyper@linkstart##1##2{}%
         \let\hyper@linkend\@empty\citet[#1][#2]{#3}}
   \biblink{#3}{\href{http://adsabs.harvard.edu/abs/#3}{ADS}}%
   \errorstopmode}            
 \newcommandtwoopt{\citeyearads}[3][][]{%
   \nonstopmode
   \href{http://adsabs.harvard.edu/abs/#3}%
        {\def\hyper@linkstart##1##2{}%
         \let\hyper@linkend\@empty\citeyear[#1][#2]{#3}}
   \biblink{#3}{\href{http://adsabs.harvard.edu/abs/#3}{ADS}}%
   \errorstopmode}            
\makeatother

\makeatletter
\newcommand{\bibnote}[2]{\@namedef{#1note}{#2}}
\newcommand{\biblink}[2]{\@namedef{#1link}{#2}}
\makeatother

\def\CaII{\ion{Ca}{II}}
\def\HI{\ion{H}{I}}

\def\dd{\ensuremath{\mathrm{d}}}
\DeclareMathAlphabet{\mathpzc}{OT1}{pzc}{m}{it}

\begin{document}

\title{Numerical non-LTE 3D radiative transfer using a multigrid method}

\subtitle{ }

\author{Johan P. Bj{\o}rgen\inst{1}
  \and Jorrit Leenaarts\inst{1} }

\offprints{J. P. Bj{\o}rgen \email{johan.bjorgen@astro.su.se}}

\institute{
Institute for Solar Physics, Department of Astronomy, Stockholm University, AlbaNova University Centre, SE-106 91 Stockholm, Sweden
}

\date{Received; Accepted }

\abstract { 3D non-LTE radiative transfer problems are computationally
  demanding, and this sets limits on the size of the problems that can
  be solved. So far Multilevel Accelerated Lambda Iteration (MALI) has
  been to the method of choice to perform high-resolution computations
  in multidimensional problems. The disadvantage of MALI is that its
  computing time scales as $\mathcal{O}(n^2)$, with $n$ the number of
  grid points. When the grid gets finer, the computational cost
  increases quadratically.} 
   {We aim to develop a 3D non-LTE radiative
  transfer code that is more efficient than MALI.} 
{ We implement a
    non-linear multigrid, fast approximation storage scheme, into the existing
    Multi3D radiative transfer code. We verify our multigrid
    implementation by comparing with MALI computations. We show that
    multigrid can be employed in realistic problems with snapshots
    from 3D radiative-MHD simulations as input atmospheres.}  
    {With multigrid, we obtain a factor 3.3-4.5 speedup compared to MALI. With
    full-multigrid the speed-up increases to a factor $6$. The
    speedup is expected to increase for input atmospheres with more
    grid points and finer grid spacing.}
     {Solving 3D non-LTE radiative transfer problems using non-linear multigrid methods can be applied to realistic atmospheres
    with a substantial speed-up. } 
    
    \maketitle


\section{Introduction}

Most of our understanding of astrophysical objects comes from analysing their spectra. The comparison of models of those objects to observations therefore requires solving the radiative transfer (RT) problem with the model as input. In this paper we investigate a non-linear multigrid method for solving the three-dimensional (3D) non-LTE RT problem in cool stellar atmospheres, and specifically, the solar atmosphere.

One-dimensional non-LTE radiative transfer has been possible for many decades. It is more recent that multidimensional radiative transfer has become possible. Now that 3D radiation-MHD simulations of solar and stellar atmospheres are widely used 
\citepads{
2003A&A...399L..31A,2004A&A...417..751A,2006ApJ...639..516U,2007ApJ...664L.135T,2009A&A...508.1403P,2013A&A...558A..20H,2013ApJ...764L..11D,2015ApJ...806...14P,2015ApJ...811...80R,
2016MNRAS.455.3735A}
it has become important to have efficient 3D non-LTE radiative transfer methods to compare these models with observations.

In the solar chromosphere, full 3D RT is important for strongly scattering lines
 \citepads[e.g.][]{2009ApJ...694L.128L,2012ApJ...749..136L,2013ApJ...772...89L,2015ApJ...803...65S,2016arXiv160605180S}.
The next solar generation telescopes, such as  the 4-meter DKIST, will obtain a diffraction limit of $~0.03''$ at 500~nm, corresponding to $32$~km on the Sun. To be able to compare these high-resolution observations with MHD simulations, 3D radiative transfer is needed even for much weaker photospheric lines 
 \citepads{2015A&A...582A.101H,2015SoPh..290..979J}.
Some important chromospheric diagnostic lines are influenced by partial redistribution effects, such as Lyman-$\alpha$, \ion{Mg}{II} h\&k, and \CaII\ H\&K. These effects have been included a 3D RT code 
\citepads{2016arXiv160605180S},
but at the cost of an increase in computational time up to a factor 10. We therefore need a robust method that can solve the 3D non-LTE problem more efficiently than currently possible.

It is well known that $\Lambda$-iteration converges prohibitively slowly in optically thick non-LTE problems. Accelerated $\Lambda$-iteration 
\citepads[ALI,][]{1973ApJ...185..621C},
converges much faster by using an approximate $\Lambda$-operator. For 3D one typically uses the diagonal of $\Lambda$-operator as the approximate operator (this choice is also known as Jacobi iteration), because it can be easily constructed without the need of computationally expensive matrix inversions.  Other methods exists: Gauss-Seidel (GS) iteration
\citepads{1995ApJ...455..646T} 
converges twice as fast as the diagonal operator, {but cannot be parallelized as efficiently as Jacobi iteration
\citepads{tsitsiklis1988comparison},
a serious disadvantage} in parallel solvers that are used for 3D applications.
Conjugate gradient methods are under development, but have not yet been implemented in a general 3D non-LTE RT code
\citepads{ 2009A&A...507.1815P}.

Therefore, multilevel accelerated lambda iteration (MALI) 
\citepads{1991A&A...245..171R,1992A&A...262..209R} 
with a diagonal approximate operator remains the current standard for 3D non-LTE radiative transfer problems for solar applications
 \citepads{2001ApJ...557..389U,2009ASPC..415...87L,2013A&A...557A.143S}.

The disadvantage with MALI is that the convergence slows down as the grid spacing gets finer
\citepads{1986JQSRT..35..431O}.
A solution to this problem was proposed by 
\citetads{1991A&A...242..290S}, 
who implemented a linear multigrid method for two-level atoms with coherent scattering in 1D and 2D test atmospheres and found a speed-up of a factor 4 to 40. The idea behind multigrid methods is that Jacobi and Gauss-Seidel iteration quickly smooth out the high spatial-frequency error in the solution, but decreases the low-frequency error only slowly. After a few iterations the error will therefore be smooth (i.e., contain only low frequencies). The problem can therefore be transferred to a coarser grid, where the low spatial-frequency errors become higher-frequency errors. Iterations on the coarse grid quickly decrease these errors. A correction to the fine-grid solution is computed on the coarse grid, and interpolated up onto the fine grid, which now has a much smaller low-frequency error.

This method was expanded by
 \citetads{1994A&A...284..319V}, 
who applied it to a 3D problem with a smooth atmosphere. He showed that the method works, but does not provide a large speed-up on small domains in parallel setup. 
\citetads{1997A&A...324..161F}
combined the non-linear multigrid technique with a multilevel Gauss-Seidel iteration scheme (MUGA). The smoothing properties of MUGA do not deteriorate with increasing grid resolution, which makes it an excellent choice to use with multigrid. They found a speed-up of a factor 3 to 32 with a five-level \CaII\ atom and a 2D atmosphere that was isothermal in the vertical direction and
contained a weak temperature inhomogeneity in the horizontal direction 
\citepads[see][Eq.~18]{1994A&A...292..599A}.
\citetads{2007A&A...470....1L} 
compared the performance of multigrid against Gauss-Seidel iteration with successive overrelaxation in smooth 2D prominence models.
\citetads{2013A&A...557A.143S}
proved that multigrid works in full 3D non-LTE with a MALI operator and parallel implementation using spatial domain decomposition. They used a plane-parallel semi-empirical FAL model C atmosphere 
\citetads{1993ApJ...406..319F},
with a sinusoidal horizontal inhomogeneity in temperature with an amplitude of 500~K.

All previous studies used relatively smooth model atmospheres. So far no one has applied non-linear multigrid for 'real life' atmospheres produced by radiation-MHD simulations which are highly inhomogeneous in all atmospheric quantities. This paper presents a non-linear multigrid method that can handle such atmospheres.

In Section~\ref{sec:method} and \ref{sec:setup} we describe the method and our computational setup . In Section~\ref{sec:numerical} we present numerical considerations for the 3D radiative transfer with multigrid. In Section~\ref{sec:results} we present the speedup we obtain compared to regular MALI iteration. Section~\ref{sec:conclusions} contains our conclusions.

\section{Method} \label{sec:method}

\subsection{The non-LTE radiative transfer problem}

The time-independent non-LTE radiative transfer problem 
consists of solving the transport equation
\begin{equation}
\frac{\dd I_{\nu}}{\dd \tau_{\nu} \left(n_i\right) } = S_{\nu}\left(n_i\right) - I_{\nu}, \label{transfer_equation}
\end{equation}
with $I_{\nu}$ the intensity, $\tau_{\nu}$ the optical path along a ray, $S_{\nu}$ the source function, and $n_i$ the atomic level populations, together with the equations of statistical equilibrium for the level populations:
\begin{equation}
 n_i \sum P_{ij}\left( I_{\nu} \right) -  \sum n_j P_{ji} \left( I_{\nu} \right)  = 0, \label{stat_eq}
\end{equation}
with $P_{ij}$ the sum of the radiative and collisional rates. We wrote the dependencies of $\tau_{\nu} $ and $S_{\nu}$ on the level populations and $P_{ij}$ on the intensity explicitly to emphasize the non-linearity of the problem. The radiation field couples various grid points together, making the problem also non-local. We assume in this paper that the collisional terms and the background opacity and emissivity are constant and fully determined by the input atmosphere. 

To close the system, we replace one of the rate equations at each grid point with a particle conservation equation:
\begin{equation}
n_{\mathrm{tot}} = \sum \limits_{i=1}^{n_{\mathrm{levels}}} n_i. \label{particle_conservation}
\end{equation}

The statistical equilibrium equation must be solved at all grid points, and the transfer equation must be solved at all grid points for all frequencies and ray propagation directions. The combination of the equations can be written as a coupled system of non-linear equations:

\begin{equation}
\mathcal{W} (n) = f. \label{eq:fundemental_eq} 
\end{equation}
We note that these are $n_\mathrm{levels} n_\mathrm{gridpoints}$ equations that in principle depend on all $n_\mathrm{levels} n_\mathrm{gridpoints}$ level populations

For a given grid point, if we replace the first rate equation with particle conservation, $f$ looks like:
\begin{equation}
    f = \begin{bmatrix}
           n_{\mathrm{tot}} \\
           0 \\
           \vdots \\
           0
         \end{bmatrix}
\end{equation}

In the MALI scheme these equations are preconditioned and solved by iteration
\citepads{1991A&A...245..171R,1992A&A...262..209R}. 
%

\begin{figure}
    \centering
      \resizebox{\hsize}{!}{\includegraphics{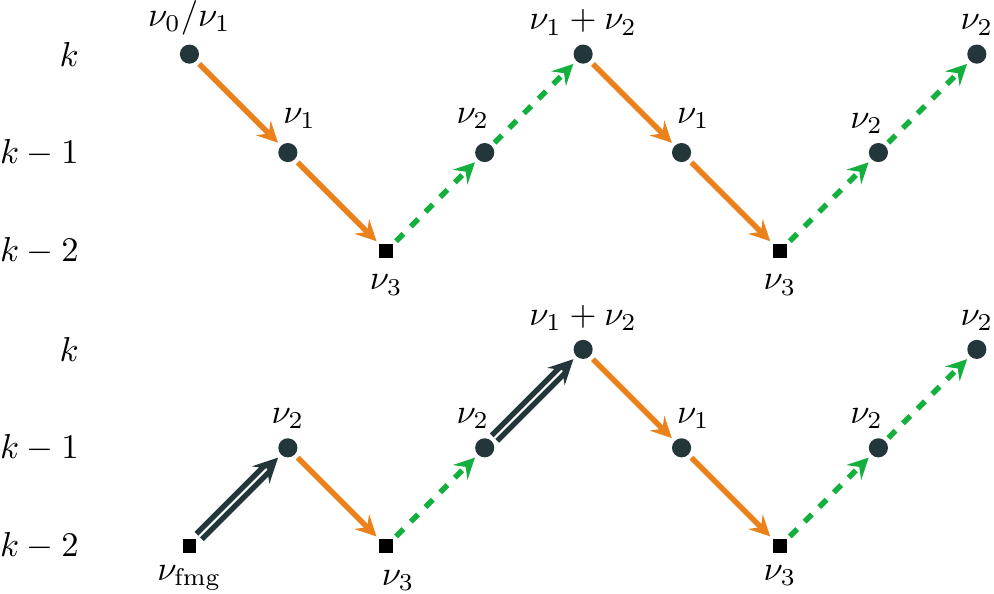}}
    \vspace{1ex}
  \caption{Schematic representation of V-cycle and full multigrid with three grid levels. $\Rightarrow$ means interpolating the full approximation, $\rightarrow$ means restricting the approximation and the residual, $\dashrightarrow$ means interpolating the correction. The iteration parameters $\nu_\mathrm{FMG}$ and $\nu_{1,2,3}$ are defined in Section~\ref{sec:method}, the parameter $\nu_0$ in Section~\ref{sec:prepostsmoothing}.}
    \label{fig:full_multigrid_illustration}
\end{figure}

\begin{figure*}[!t]
  \begin{minipage}{\textwidth}
    \includegraphics[width=0.5\textwidth]{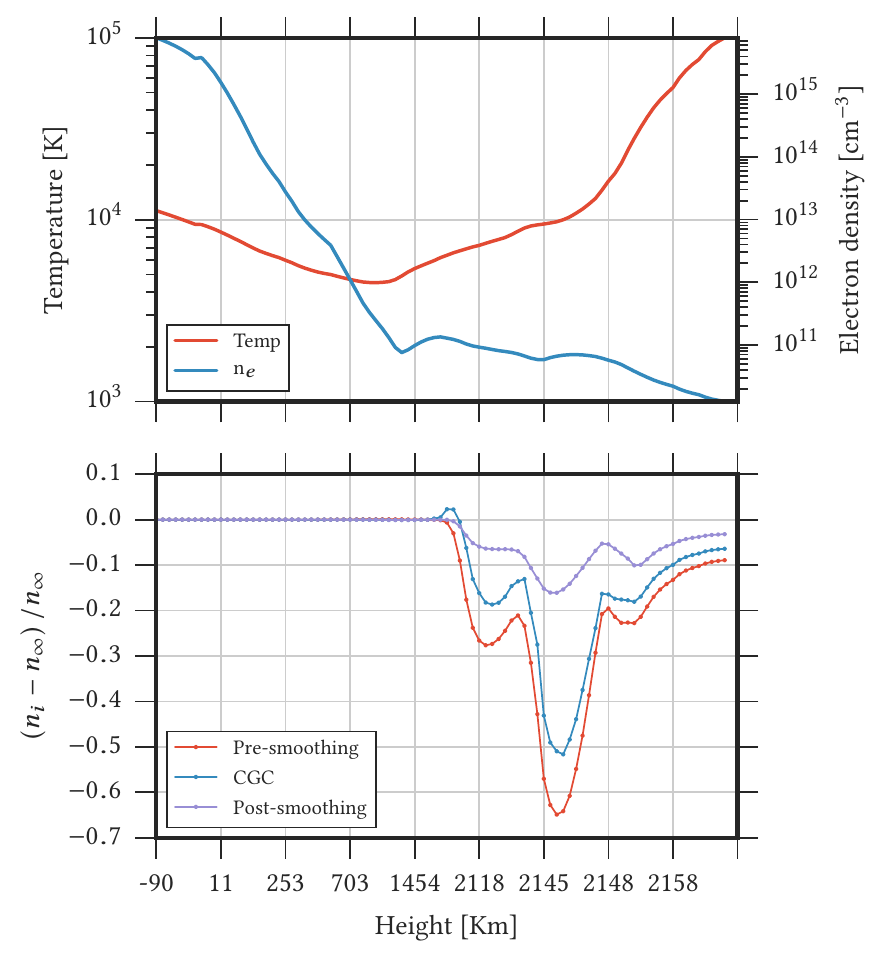}\hfil
    \includegraphics[width=0.5\textwidth]{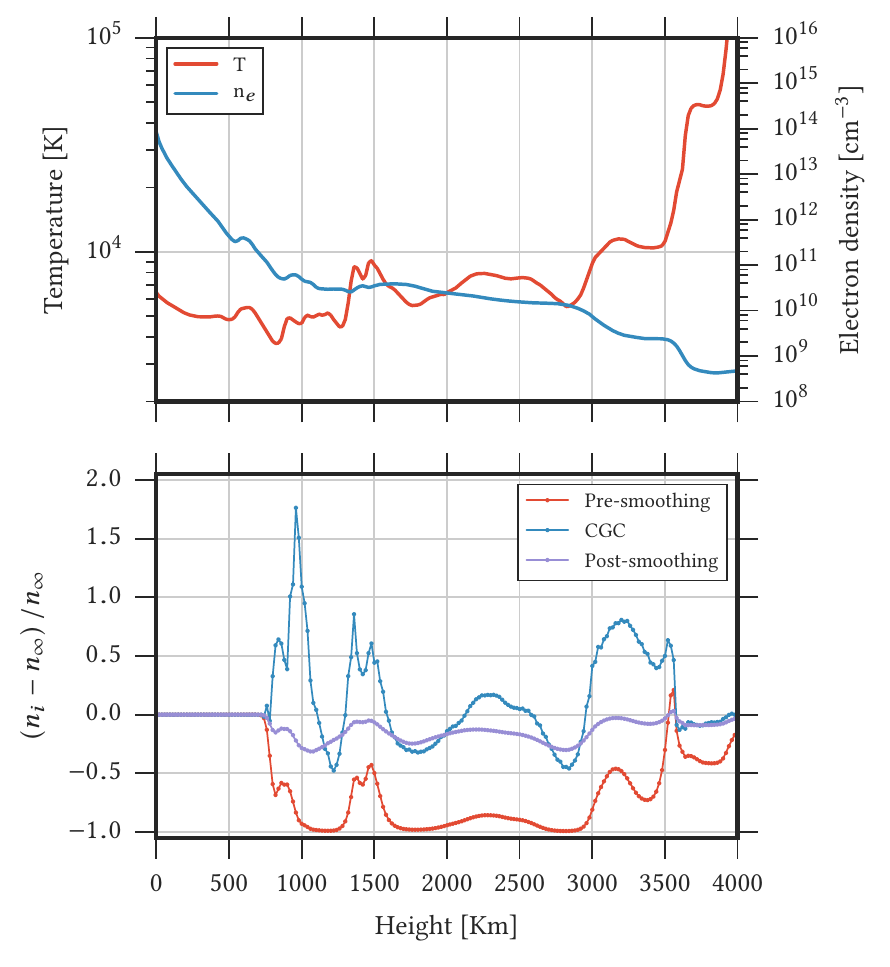}\hfil
  \end{minipage}
  \caption{Behaviour of the relative error in population at the finest grid during various steps in a V-cycle with three grids, for the FAL-C atmosphere (left-hand panels) and a column extracted from a 3D radiation-MHD atmosphere (atmosphere Model 1, see Section~\ref{sec:setup}). Both computations are done for an \HI\ atom using $10$ pre-smoothings, $32$ coarse-grid iterations and $25$ post-smoothings. The upper panel shows the temperature and electron density in the atmosphere. The lower panels show the error for the \HI\ $n=2$ level after the pre-smoothing (red), the error after application of the coarse-grid correction (blue) and the error after application of the post-smoothing (purple). For both atmospheres the coarse grid correction has decreased the low spatial-frequency error, but increased the high-frequency error. The latter is largely removed by the post-smoothing.  
  \label{spectra_radius_smooth_atm}}
\end{figure*}

\subsection{The non-linear multigrid method} \label{sec:multigrid}

In this subsection we present how the non-linear multigrid method is applied to the radiative transfer problem, i.e. how it is used to solve  Eq.~\ref{eq:fundemental_eq}.

The non-linear multigrid, fast approximation storage (FAS) scheme as formulated by
 \citetads{brandt1977multi},
 preserves the mathematical structure of $\mathcal{W}$ at the coarser grids. For non-LTE codes this means that one can reuse the formal solver, computation of collisional rates, etc., and that the iterative MALI method can be used on the coarser grids with only minor modifications. 

We restate the rate equation:
\begin{equation}
\mathcal{W}^{k} (n^k) = f^k, 
\end{equation}
where $k$ is an index that denotes the grid level: $k=\ell,\ell-1, \dots,1$ where $\ell$ is the number of grids, with $k=\ell$ the finest grid and $k=1$ the coarsest grid. We provide an initial guess for $n$ and perform a number ($\nu_1$) MALI iterations, which is called pre-smoothing. Because MALI iterations act as a smoothing operator, we have reduced the high-frequency error and obtained a smooth residual $r^{k}$:
\begin{equation}
r^{k} =  f^{k} - \mathcal{W}^{k}_{*} (n^{k}) \label{residual_equation}.
\end{equation}

The $*$, denotes that an extra formal solution has been performed to get the radiative rates consistent with the current population estimate. 

The exact solution can be described by adding a correction to the approximation: $n^k_{\rm{exact}} = n^k + e^k $. The exact fine grid equation is then:
\begin{equation}
\mathcal{W}^{k}  (n^{k} + e^{k}) = f^{k}. \label{fine_grid_exact}
\end{equation}

Using the residual equation (Eq. \ref{residual_equation}) the fine grid equation (Eq. \ref{fine_grid_exact}) may be written as

\begin{equation}
\mathcal{W}^{k}  (n^{k} + e^{k}) = r^{k} + \mathcal{W}^{k}_{{*}} (n^{k}).
\end{equation}
All these quantities can be mapped to a coarser grid $k-1$ using a restriction operator $ \mathcal{R}^{k-1}_{k}$. We point out that we assume that only the residual is smooth and not the populations themselves. The populations from the fine grid are used as an initial approximation at the coarse grid. The coarse grid operator $\mathcal{W}^{k-1}$ is obtained by performing a formal solution at the coarse grid. This results in the coarse-grid-equation:

\begin{equation}
\mathcal{W}^{k-1}  (n^{k-1} + e^{k-1}) =  \mathcal{R}^{k-1}_{k} ( r^{k} ) + \mathcal{W}^{k-1}_{{*}}   \left( \mathcal{R}^{k-1}_{k}  (n^{k}) \right). \label{coarse_eq}
\end{equation}

The right-hand-side tries to force the solution of the system to maintain the fine grid accuracy. 
The coarse grid equation does not need to be solved exactly, since it is limited by the accuracy of the right-hand-side. Instead we compute an approximate solution by applying a number $\nu_3$ MALI iterations, called coarse grid iterations. The relative correction (See Section~\ref{correction}) is then computed by:
\begin{equation}
e^{k-1} = \frac{ n_{\mathrm{after}}^{k-1} - n^{k-1}_{\mathrm{before}}}{n^{k-1}_{\mathrm{before}}}, \label{eq:relcor}
\end{equation}
where the subscripts before and after denote the populations before and after the $\nu_3$ MALI iterations.

The correction is mapped onto the operator to the finer grid with an interpolation operator $\mathcal{I}_{k-1}^{k}$:

\begin{equation}
n^{k}_\mathrm{new}  = n^{k}_\mathrm{old} \left( 1 + \mathcal{I}_{k-1}^{k}  ( e^{k-1} ) \right) \label{eq:interpolaterelcor}
\end{equation}

The interpolation introduces high-frequency errors on the fine grid. By applying $\nu_2$ MALI iterations on the fine grid (called post-smoothing), these high-frequency errors are removed, and we have obtained a better approximation to the true solution on the finest grid. To obtain a converged solution, multiple cycles have to be performed.  

The method we just described is called two-grid FAS. The same procedure can be applied to Eq.~\ref{coarse_eq} because it has the same structure as Eq.~\ref{eq:fundemental_eq}. Doing this leads to the general multigrid procedure. The successive coarsening starting from the finest grid and the interpolation steps from the coarsest to the finest grid together are called a V-cycle.

\subsection{Full multigrid}

Full multigrid (FMG) can be seen as a method to obtain a better initial approximation at the finest level. FMG is considered to be the optimal multigrid solver
 \citepads{multigrid_trottenberk}.
We explain FMG using a three-grid ($\ell=3$)  example, as illustrated in Figure~\ref{fig:full_multigrid_illustration}. In FMG, Eq.~\ref{eq:fundemental_eq} is solved at the coarsest grid:

\begin{equation}
\mathcal{W}^{k=1} \left( n^{k=1} \right) = f^{k=1}.
\end{equation}

The initial population and the right-hand-side can be obtained by restricting it from the finest grid or direct initialization at the $k=1$ grid. After applying a $\nu_\mathrm{FMG}$ iterations, an initial approximation is acquired at the coarsest level. The full approximation is then interpolated up to the next finer grid $k=2$:
\begin{equation}
n^{k=2}  =  \mathcal{I}_{k=1}^{k=2}  \left( n^{k=1} \right).
\end{equation}

The new approximation is used as the initial population for solving Eq.~\ref{eq:fundemental_eq} at the $k=2$ grid. Now a V-cycle is applied at this level, instead of interpolating up to the finest grid $k=3$. The reason for this is that the approximation at $k=2$ will have both low-frequency and high-frequency errors due to the interpolation and discretization errors. These errors are efficiently reduced by a V-cycle. When the V-cycle is completed, we interpolate the population at $k=2$ to the finest grid $k=3$. Now we have an initial approximation at $k=3$ that is closer to the true solution than a straight initialization at grid $k=3$.

Remember that we are not aiming to get the full solution for the radiative transfer problem, but rather a better initial approximation at the finest grid. After reaching the finest level, the FMG is followed by regular FAS V-cycles to obtain the solution.

In Figure~\ref{fig:full_multigrid_illustration} we show a schematic representation of the FAS and full-multigrid-algorithms. Figure~\ref{spectra_radius_smooth_atm} shows an illustration of the behaviour of the relative error in a smooth semi-empirical atmosphere and a non-smooth column from a radiation-MHD simulation, but with otherwise identical setup. Comparing the two cases shows that the radiation-MHD simulation column shows much more high-frequency error after the coarse grid correction than in FALC. These high frequency errors stem from the inevitable inability to accurately represent a non-smooth atmosphere on a much coarser grid. We will show in this paper that multigrid can still handle such atmospheres, but at the cost of an increase in computational time and thus a smaller increase in computational speed compared to MALI than reported in earlier works for smooth atmospheres.

\section{Computational setup} \label{sec:setup}

We implemented the FAS multigrid method including FMG into the existing radiative transfer code Multi3D
 \citepads{2009ASPC..415...87L}.
 This code solves the statistical equilibrium equations for one atomic species  with a background continuum using MALI with a diagonal approximate operator  
 \citepads{1991A&A...245..171R,1992A&A...262..209R}. 
The code used 3D atmospheres on a Cartesian grid as input.
It is parallelized with MPI\footnote{Message Passing Interface} using spatial domain decomposition as well as decomposition over frequency. The atomic bound-bound transitions can be treated in complete or partial redistribution
 \citepads{2016arXiv160605180S}. 
The  transport equation is solved using short-characteristic
 \citepads{1988JQSRT..39...67K} 
 with a linear or Hermite-spline 
 \citepads{2003ASPC..288....3A,2013A&A...549A.126I}
 interpolation scheme. We use the 24-angle quadrature (A4 set) from 
 \citetads{carlson1963methods}. 
 All computations presented in this paper are done assuming complete redistribution and without frequency parallelization.

\subsection{Model atmospheres}

We use three different model atmospheres in this paper.

The first is a snapshot from the radiation-MHD simulation by 
\citetads{2016A&A...585A...4C}
computed with the Bifrost code
\citepads{2011A&A...531A.154G}
taken at t = 3850 s

This model extends from the upper convection zone to the corona in a box with $504\times 504\times 496$ grid points, with a horizontal grid spacing of 48 km, and a vertical grid spacing ranging from 19 to 100 km. The model contains strong gradients in temperature, velocity and density and has been used before in several studies
 \citepads[e.g.][]{2013ApJ...764L..11D,2013ApJ...772...89L,2013ApJ...772...90L,2015ApJ...803...65S,2015ApJ...806...14P,2015A&A...575A..15L,2015ApJ...811...80R,2015ApJ...811...81R} . 
We used this model to see how multigrid behaves in a realistic use case. To obtain an odd number of grid points in the vertical direction (see Section~\ref{sub_domain}), we added one extra layer in the convection zone through linear extrapolation of all quantities, so we obtain $504\times 504\times 497$ grid points. This MHD snapshot will be called atmosphere Model 1.

We also use a Bifrost simulation snapshot with $768\times 768\times 768$ grid points (Carlsson \& Hansteen, private communication).  In this snapshot the $x$- and $y$-axis are equidistant with a grid spacing of $32$ km. The vertical axis is non-equidistant, with a grid spacing $13$ km in the photosphere and the chromosphere, and increasing to 60~km in the convection zone and the corona. The physical extent is the same as for Model 1: $24 \ \mathrm{Mm} \times 24 \ \mathrm{Mm} \times 16.9 \ \mathrm{Mm}$. We only used the part of the atmosphere between $z_{\mathrm{min}}=-0.5 \ \mathrm{Mm}$ and $z_{\mathrm{max}}=6 \ \mathrm{Mm}$, to cover the formation height of the \CaII\ lines. The grid size is therefore reduced to $768\times 768\times 473$. This simulation has a similar magnetic field configuration as Model~1, but was computed with an LTE equation of state instead of the non-equilibrium equation of state of Model 1. This atmosphere will be called Model $2$.
Finally we also use some computations using the plane-parallel semi-empirical FAL model C atmosphere
\citepads{1993ApJ...406..319F}, 
which we will refer to as FAL-C.

\subsection{Model atoms}

We use three different model atoms. Most test computations were done with a minimalistic three-level \CaII\ atom, containing the ground level, the upper level of the  \CaII~K line and the \ion{Ca}{III} ground state. We also use the 6-level \CaII\ atom and the six-level hydrogen atom from
\citetads{2012A&A...539A..39C}.
All bound-bound transitions are treated in complete redistribution.

\subsection{Convergence criteria and error estimates}

To validate and compare our multigrid implementation with the standard one-grid MALI scheme, we compare against reference solutions computed using the one-grid MALI scheme.  With the three-level \CaII\ atom we converged the reference solution  to $\rm{max} \left| \delta n/n \right| \leq 10^{-5}$.  For the hydrogen atom we converged to $\rm{max} \left| \delta n/n \right| \leq 10^{-4}$. The reason for the lower criterium for hydrogen was to limit computational expenses: to reach the limit for hydrogen we needed roughly $300,000$ CPU hours.

To characterize the absolute error after $i$ iterations we used the infinity norm:
\begin{equation}
E_{i,\infty} = \max \left| \frac{ n_{i}-n_{\infty}}{n_{\infty}} \right|,
\end{equation}
where the $n_{\infty}$ populations are taken to be the reference solutions.
Since one grid point can slow down or stall the convergence with multigrid, the infinity-norm is the best choice. Other norms, such as the 2-norm, can give a false picture of the convergence of the multigrid method.  

The maximum relative correction in population during each iteration is given by
\begin{equation}
C_i = \max \left| \frac{ n_i-n_{i-1}}{n_i} \right|.
\end{equation}
For linear non-LTE radiative transfer problems one can define the spectral radius as the largest eigenvalue of the amplification matrix of the iteration scheme. However, in the non-linear case, this is not possible. Therefore we define the spectral radius operationally as  
\begin{equation}
 \rho_{\mathrm{sr}} = \frac{E_{i+1}}{E_i}. \label{eq:spectral_radius_def}
\end{equation}
It measures how much the error is reduced for each iteration.

For normal use cases there is no reference solution, so the error cannot be computed directly. For MALI iterations it was shown 
\citepads{1994A&A...292..599A,1997A&A...324..161F}
that one can use the maximum relative correction to estimate the error. When the relative correction reaches the asymptotic linear regime, the relation between the error, the spectral radius and the relative correction is:
\begin{equation}
E_i \approx  \frac{\rho_{\mathrm{sr}}}{1-\rho_{\mathrm{sr}}} C_i. \label{eq:error_approx}
\end{equation}
Because there is no reference solution, the spectral radius is estimated using  $\rho_\mathrm{sr} \approx C_{i+1}/C_i$. We test whether this approximation is also valid for multigrid iteration in Sec~\ref{sec:spandconv}.

\section{Numerical considerations} \label{sec:numerical}

In the following subsections we discuss a number of issues that are relevant for radiative transfer using multigrid methods.

\subsection{Grid coarsening}

We use a simple coarsening strategy {for the grid coordinates}, by removing every second grid point along the $x$, $y$ and $z$ axes when coarsening from grid $k$ to $k-1$.

\subsection{Grid sizes}

In the vertical direction we use fixed non-periodic boundaries. In the lower boundary, the incoming radiation is set to the Planck function at the local gas temperature. In the upper boundary, the incoming radiation is set to zero or to pre-specified values. We chose to keep the vertical boundaries at fixed heights. To obtain this, we require an odd number of grid points in the vertical direction at all grid levels. Therefore the number of vertical grid points $N_Z$ at the finest level must fulfil the following relation:
\begin{equation}
N_{z} = A \cdot 2^{\ell}+1,
\end{equation}
where $A$ is an integer and $\ell$ the number of grids.

In the horizontal plane we use periodic boundary conditions. Because Multi3D requires constant grid spacing in the horizontal direction
we thus require that $N_{x,y}$, the number of grid points in the $x$ or $y$ direction, follows
\begin{equation}
N_{x,y} = B \cdot 2^{\ell},
\end{equation}
with $B$ an integer.

\subsection{Sub-domain sizes and number of grids} \label{sub_domain}

\begin{figure}
    \centering
      \resizebox{\hsize}{!}{\includegraphics{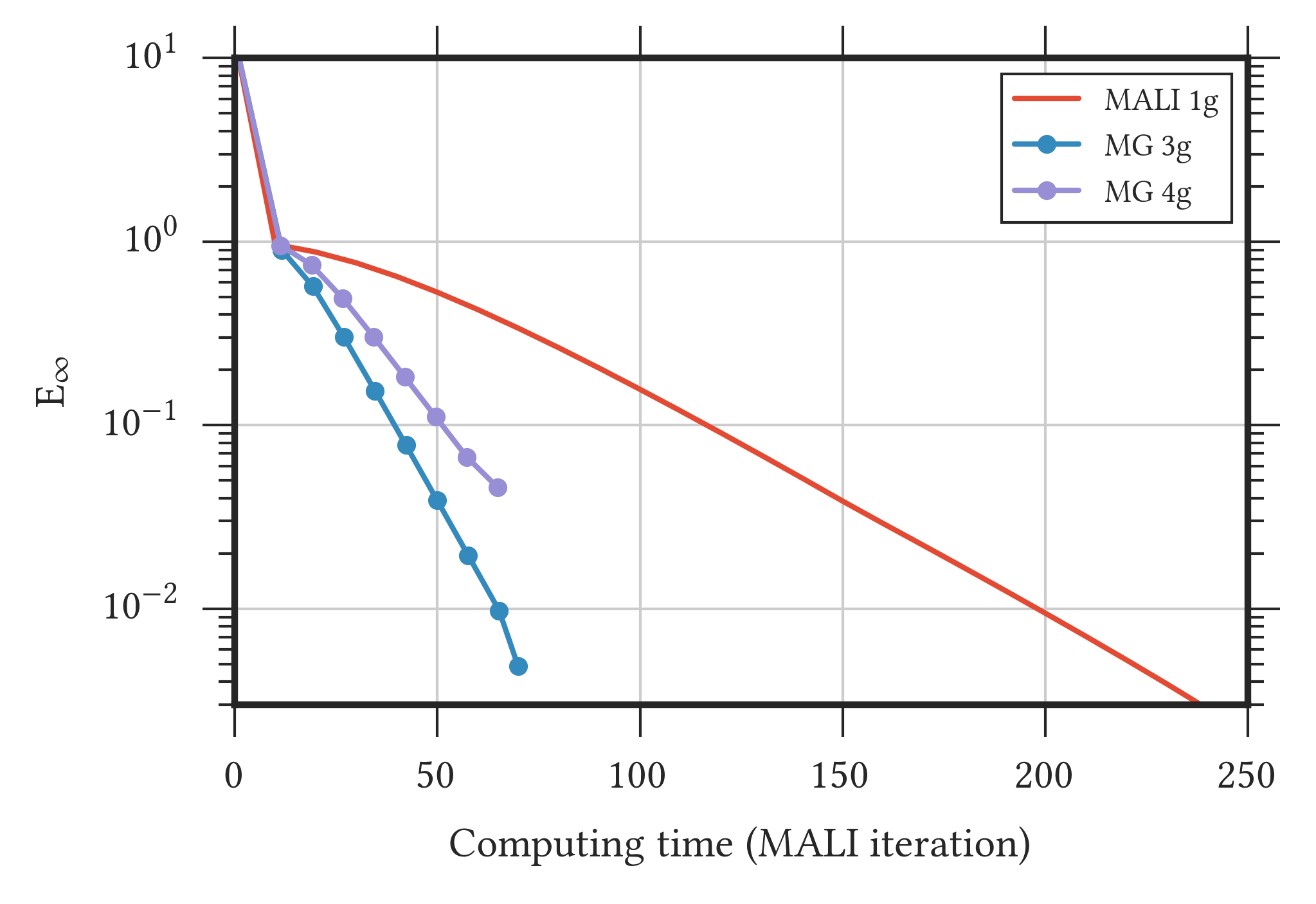}}
    \vspace{1ex}
  \caption{Convergence behavior for MALI (red) and multigrid with three (blue) and four (purple) grids for the three-level \CaII\ atom with in atmosphere Model 1. {The computing time is expressed in units of the time it takes to perform one MALI iteration at the finest grid.} The multigrid computations were performed with $\nu_1 = 2$, $\nu_2 = 2$, $\nu_3=32$, full-weighting restriction and trilinear interpolation.}
    \label{fig:3dbifrost_3g_4g_caii}
\end{figure}

Multi3D divides the computational domain into sub-domains. The formal solver requires 5 ghost zones in the horizontal directions, and the number of internal points in the horizontal direction must be at least as large as the number of ghost zones. In the vertical direction we have only one ghost zone. In single-grid applications, the subdomains have typical sizes from $32^3$ to $64^3$. In multigrid, we use the same domain decomposition as in the finest grid, and coarsen the grid local to each sub-domain. This limits the number of coarsenings that we can do.  For a $32^3$ subdomain, we can achieve two coarsening to $16^3$ and $8^3$. A $64^3$ subdomain can be coarsened 3 times. Through various tests we found that three grids gives us the best and stable performance. Using more than three grids can lead to non-convergence or lower performance than using three grids
\citepads[see also Table 1 of][]{1991A&A...242..290S}.
In Figure \ref{fig:3dbifrost_3g_4g_caii} we display the result of such a test for the three-level \CaII\ atom and atmosphere Model 1.

\subsection{Restriction operator} \label{restriction_operator}

\begin{figure}
    \centering
      \resizebox{\hsize}{!}{\includegraphics{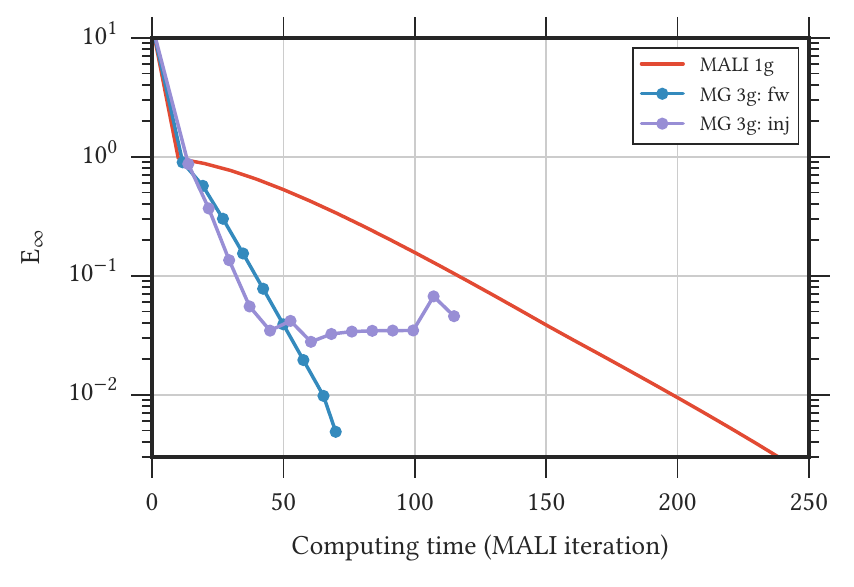}}
    \vspace{1ex}
  \caption{Convergence behavior for MALI (red) and multigrid with injection (purple) and full-weighting (blue) as restriction operators, using the three-level \CaII\ atom and atmosphere Model 1. The multigrid computation was performed with $\nu_1 = 2$, $\nu_2 = 2$, $\nu_3=32$, three grids and trilinear interpolation. 
 }
    \label{fig:3dbifrost_inj_fw_caii}
\end{figure}

We implemented three different restriction operators: injection, half-weighting and full-weighting 
\citep[e.g.][]{multigrid_trottenberk}. 
Injection is the most straightforward method, as it just removes grid points when moving a quantity from a fine to a coarse grid. Half-weighting is a smoothing operation, where a coarse grid point value is an average of the same point plus its six neighbour grid points located along the $x$,$y$, and $z$ axes on the fine grid.  Full-weighting is similar, but includes all 26 neighbour grid points. Full-weighting has the advantage that it conserves a quantity when it is integrated over the entire computational domain. 
Injection is computationally more efficient than the half and full weighting methods, but the extra computing time is negligible compared to the time spent on the formal solution. 

We performed various test using both the FAL-C and Model 1 atmospheres. We found that injection performs well for the FAL-C atmosphere, but not for Model 1. Figure \ref{fig:3dbifrost_inj_fw_caii} show a comparison between single-grid MALI, three-grid injection, and three-grid full-weighting using the three-level \CaII\ atom in Model 1. The convergence of multigrid using injection stalls around $E_\infty = 0.03$, while the full-weighting computation converges to the reference solution.
We therefore chose to use full-weighting {for the atomic level populations and residuals} for all further computations.

\subsection{Interpolation operator}

We implemented two interpolation operators: linear and cubic Hermite 
 \citepads{2003ASPC..288....3A,2013A&A...549A.126I}.
The coefficients for linear interpolation depend only on the grid spacing and can be precomputed.  A full 3D interpolation uses information of the 8 surrounding grid points. Cubic Hermite interpolation needs a $4\times4\times4$ stencil for a 3D interpolation and requires the computation of the derivatives of the quantity to be interpolated on the inner 8 grid points. This makes cubic Hermite interpolation somewhat more involved to implement and more computationally expensive. We performed a number of 1D test calculation using columns from Model 1 and found that, on average, both interpolation methods perform equally well. We therefore chose to use the simpler and faster linear interpolation for our 3D computations.

\subsection{Correction} \label{correction}

The radiative transfer problem is highly non-linear, and a level population can in principle change by orders of magnitude from one grid point to the next. This poses a problem for the interpolation of the correction from a coarser to a finer grid (Eqs.~\ref{eq:relcor} and~\ref{eq:interpolaterelcor}) because the interpolation operation can introduce a large amount of interpolation noise and even lead to  the unphysical prediction of negative populations at the fine grid. Both cases destroy the good convergence behaviour of multigrid. We found that interpolating the relative correction (Eq.~\ref{eq:relcor}) decreases the interpolation noise and largely eliminates the occurrence of negative populations as a consequence of interpolation.

\subsection{The atmosphere at various grid levels}

Multigrid requires solving the formal solution at all grid levels. We thus need a method to generate coarse-grid model atmospheres from the fine-grid atmosphere. We tested both injection and full weighting to do so. Our tests indicate that injection can lead to the best convergence rate, especially in smooth atmospheres like FAL-C, but it might also cause stalls in convergence or even divergence. Full-weighting has a lower best-case convergence, but does not lead to stalls or divergence. The reason for this difference is that injection more accurately represents the fine-grid problem, and so has better best-case performance. But it also enhances gradients on the coarser grid, leading to stronger non-linearities and larger probability of stalling. Full-weighting on the other hand leads to  coarse-grid atmospheres that are more different from the fine-grid atmosphere, leading to lower performance, but also has smaller gradients leading to increased stability.
Because stability is usually the greatest concern when using radiation-MHD atmospheres,  we use full-weighting {for the atmosphere initialization} in our computations.

\subsection{Initializing the radiation field at various grid levels}

Multi3D treats background processes using LTE opacities and a source function with a thermal part and a coherent-scattering part. At each grid level, we thus need to perform a small number of $\Lambda$-iterations to ensure the background processes are initialized correctly. Because a $\Lambda$-iteration takes a similar time as a MALI iteration we want to minimize the number of $\Lambda$-iterations. Empirically we found that one  $\Lambda$-iteration is sufficient at the finest grid. At the coarser grids we obtain an initial guess for the radiation field by restricting it from the next finer grid and performing one or two $\Lambda$-iterations.

\subsection{Computational cost}

\begin{figure}
  \resizebox{\hsize}{!}{\includegraphics{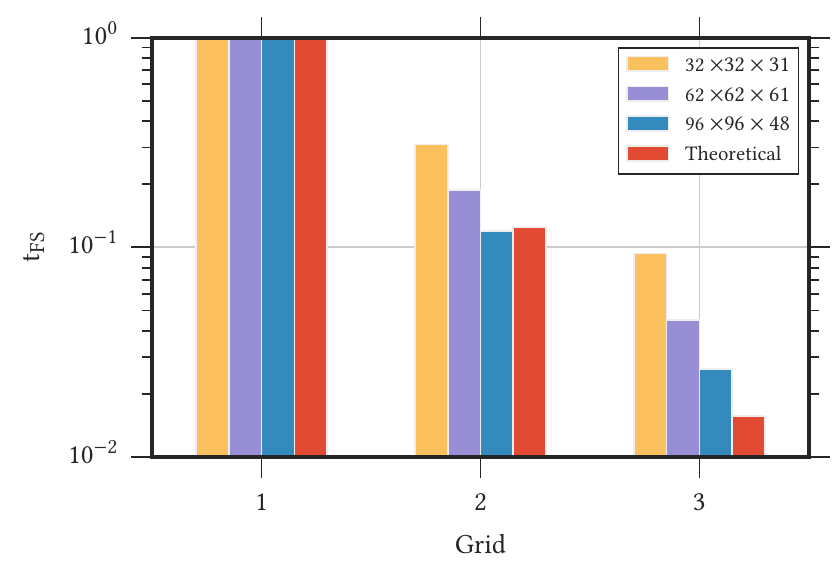}}
  \caption{
  The computational cost for one formal solution at each grid in multigrid with different sub-domain sizes. We set the cost of a formal solution at the finest grid at 1. The theoretical value is calculated as $(1/8)^{\ell-k}$, with $k$ the grid level.}
  \label{speed_up_multigrid}
\end{figure}

In this paper we are mainly concerned with the theoretical performance of multigrid compared to MALI iteration. We therefore express computation time in units of one MALI iteration at the finest grid in our figures. The computing time {of a formal solution} at coarse grid $k-1$ is then $1/8$ times the computing time at grid $k$. 

However, Multi3D does not follow this theoretical perfect scaling. The reason is that the short-characteristic solver in Multi3D requires 5 ghost zones in the horizontal direction and one ghost zone in the vertical direction. As the number of grid points per subdomain gets smaller with increasing coarsening, the relative amount of time spent in updating the ghost zones increases. In Fig.~\ref{speed_up_multigrid} we compare the theoretical maximum speed-up with the actual speed up measured on a HP Cluster Platform 3000 with Intel Xeon E-5 2600 2.2 GHz cores using three-grids. At grid $k=2$ the performance penalty is a factor 3 for the $32\times32\times31$ subdomain, but interestingly the $96\times96\times48$ subdomain shows a speedup better than the theoretical value. At the $k=1$ grid the performance penalty is visible for all subdomain sizes that we tested, but with the expected behaviour that the larger the subdomain, the smaller the performance penalty. 

\subsection{Occurrence of negative populations}

\begin{figure*}
  \begin{minipage}{\textwidth}
        \includegraphics[width=\textwidth]{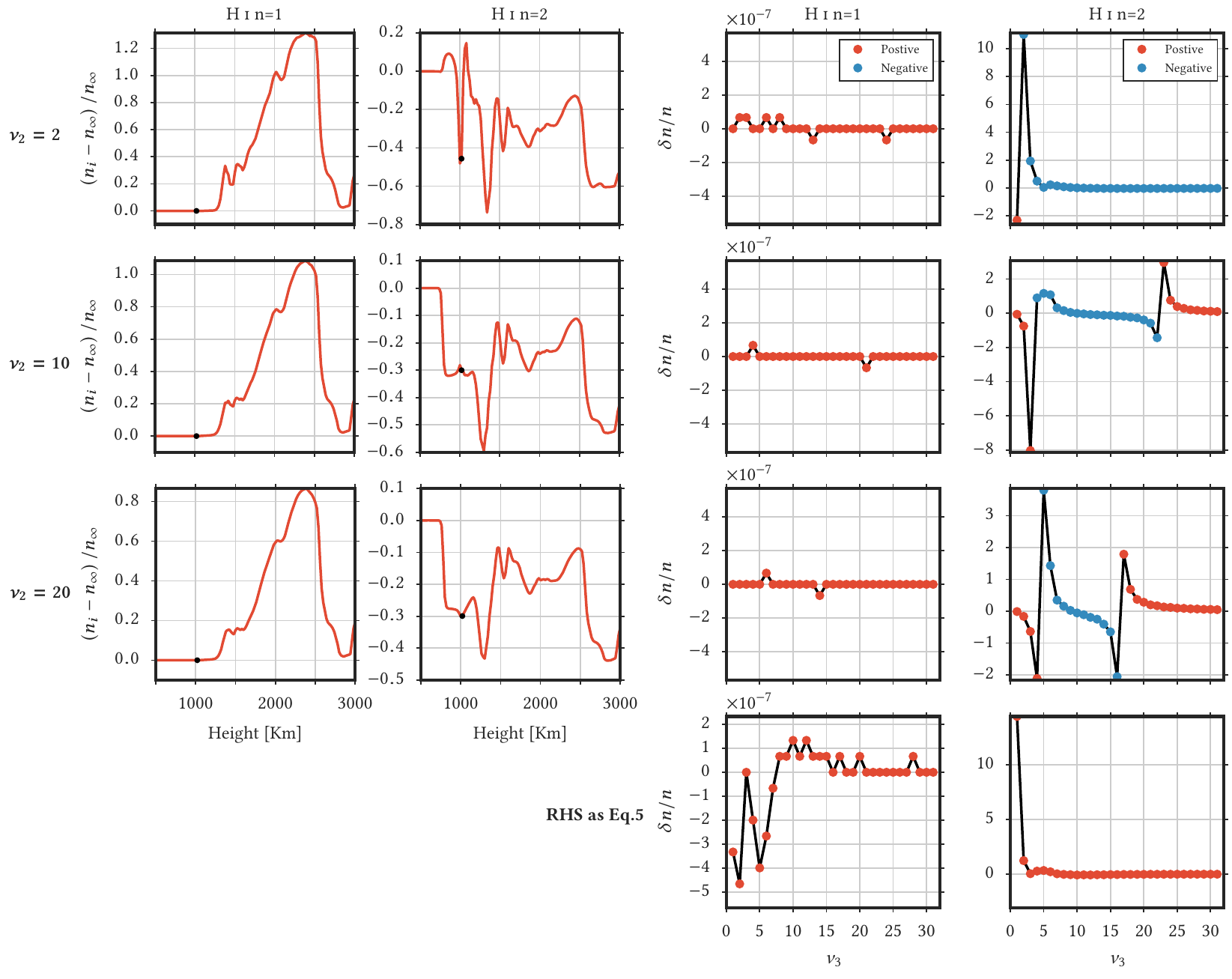}\hfil
  \end{minipage}
  \caption{Dependence on the number of post-smoothing iterations of the occurrence of negative populations at the coarsest grid. The left-hand columns show the relative error at the finest grid after two full V-cycles.
 The right-hand columns show the relative change of the population at the coarsest grid in a single grid point in the atmosphere. Red dots indicate positive populations; blue dots negative populations. The chosen grid point is indicated by the black dot in the left-hand columns. 
For each row of panels a different number of post-smoothing iterations is applied. First row: $\nu_2 = 2$; second row: $\nu_2 = 10$; third row: $\nu_2 = 20$.
In the fourth row we show the relative change of the population at the coarsest grid with $\nu_2 = 2$, but we solve Eq.~\ref{eq:fundemental_eq} instead of Eq.~\ref{coarse_eq}, i.e., we perform standard MALI iterations instead of coarse-grid iterations.
The computation was performed in a 1D plane-parallel atmosphere constructed from column $x=0$, $y=244$ of Model 1, with the \HI\  atom. The other multigrid parameters are $\nu_1=2$, $\nu_3=32$, full-weighting restriction and linear interpolation.  }
    \label{fig:negative_pop_coarse_grid}
\end{figure*}

Standard MALI iteration keeps atomic level populations positive as long as the iteration is started with positive populations. This is however not true for multigrid iterations. The reason is that coarse grid iterations have a modified right-hand side in the rate equations:
\begin{equation}
\resizebox{1.\hsize}{!}{$\mathcal{W}^k(n^k)=\begin{cases}f^{k},&\text{if }k=\ell\\\mathcal{R}^{k}_{k+1}(f^{k+1}-\mathcal{W}^{k+1}(n^{k+1}))+\mathcal{W}^{k}(\mathcal{R}^{k}_{k+1}  (n^{k+1})),&\mathrm{if }k<\ell\end{cases}$}
\end{equation}
The equation for $k<\ell$ can in principle have a negative population as a solution, or produce negative estimates of the solution during iteration. Negative populations on the coarse grids will propagate to the finest grid and destroy the convergence. 

By experimentation we found that negative populations at the coarse grid typically occur when the error $(n_i-n_\infty)/n_\infty$ is not smooth enough, but we did not manage to get an unambiguous definition of what is 'smooth enough'. The error is sometimes not smooth at the location of strong gradients in the atmospheric parameters, but sometimes large error fluctuations also occur in smooth parts of the atmosphere models. The error is typically less smooth in our \HI\ atom than in the  \CaII\ atoms. This is caused by the larger spectral radius for hydrogen (see Section~\ref{sec:spandconv}), which lowers the smoothing capability of Jacobi iterations.

Figure \ref{fig:negative_pop_coarse_grid} illustrates this in an $\ell=3$ test computation using a non-smooth 1D atmosphere and the \HI\ atom. The left-hand columns show the error for the first two energy levels of the atom after two V-cycles but with a different number of post-smoothing iterations. The error for the $n=2$ level exhibits a strong spike at a height of 1000~km for $\nu_2=2$, but this spike is much smaller for $\nu_2=10$ and $\nu_2=20$. In contrast, the error in the $n=1$ population is relatively smooth, and at 1000~km height it is very small. 

The black dot in the left-hand panels indicates the grid point that we investigate in more detail in the right-hand panels. The right-hand panels show the relative change in the $n=1$ and $n=2$ populations as a function of iteration at the coarsest grid ($k=1$). The corrections are small for $n=1$, and the $n=1$ population remains positive for all coarse-grid iterations, as expected from the small error seen in the left-hand panels.

The $n=2$ population behaves remarkably different. For $\nu_2=2$, the first coarse-grid iteration predicts a relative change of -1.7 and thus produces a negative population. Successive corrections are smaller but the population remains negative. When $\nu_2=10$ (second row), the population is positive for the first 3 iterations, but then turns negative until iteration 22, after which it is positive again. For $\nu_2=20$ the behaviour is similar, but the number of iterations where the populations are negative is shorter.

For comparison, in the fourth row we show the relative change at the coarsest grid with $\nu_2=2$ but where we perform standard MALI iterations (Eq.~\ref{eq:fundemental_eq}) instead of coarse-grid iterations (Eq.~\ref{coarse_eq}). Here the populations are always positive, demonstrating again that MALI iterations keeps a positive solution positive.

If negative populations occur after application of the correction (Eq.~\ref{eq:relcor}), we simply replace it with $10\%$ of the  uncorrected population and continue as usual with post-smoothing iterations. If this happens at isolated points in the atmosphere then the convergence of the multigrid iteration is not affected, because a single outlying grid-point represents high-frequency noise, which is damped away efficiently.  However, if there are too many negative populations in the same area of the atmosphere then convergence will be slower or the iteration might even diverge. 

\subsection{Initial population}

\begin{figure}
  \resizebox{\hsize}{!}{\includegraphics{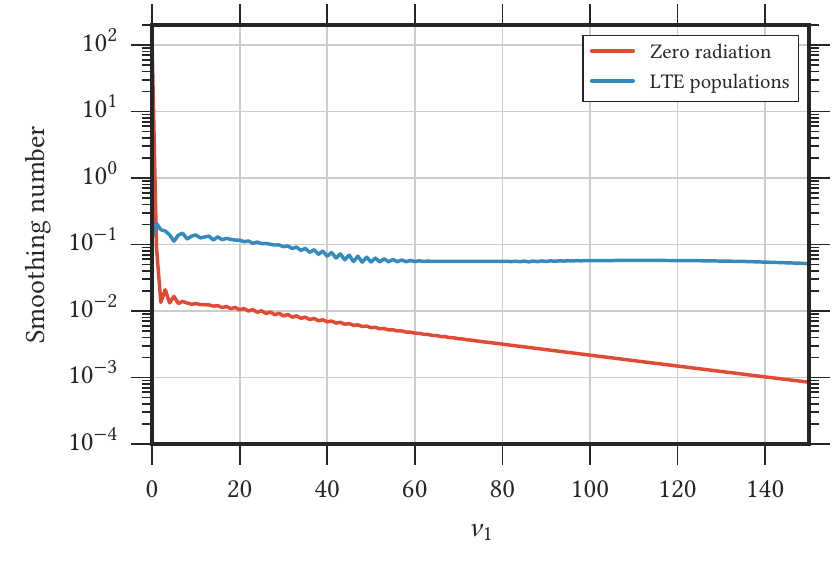}}
  \caption{Dependence of the smoothing number $\rho_L(\nu_1)$ ({ see Eq. \ref{eq:smooth_eq}}), on the initial condition of the populations. The results were computed using column $x=0,y=271$ from Model 1 as a one-dimensional atmosphere and the hydrogen model atom.}
  \label{smoothing_number}
\end{figure}

A good initial approximation for the level populations is critical to for good convergence. There are two popular methods to initialize the populations: LTE and the zero-radiation-field approximation. The latter means solving the statistical equilibrium equations with the radiation field set to zero everywhere in the  atmosphere. We found that LTE-populations as an initial approximation typically leads to negative populations at the coarse grid. 

We suspect that the reason is that the residual and the error are less smooth using LTE initialization than using the zero-radiation approximation. To obtain a smooth residual and error many more iterations are required. We demonstrate our conjecture by comparing  the smoothing number \citep[]{hackbush,1997A&A...324..161F} as function of MALI iterations at the fine grid for both initializations.
The smoothing number is defined as

\begin{equation}
\rho_L (i) = max \left|  { \mathcal{D}_2 \left[ n_{i}-n_{\infty}  \right] } \right|, \label{eq:smooth_eq}
\end{equation}
where the operator $\mathcal{D}_2$ is the second-order derivate and $i$ is number of MALI iterations. In Figure \ref{smoothing_number}, we show that the zero-radiation initialization leads to a much smoother error than LTE initialization. A similar result is obtained for the residual. We therefore use zero-radiation initialization for all our computations.

\subsection{Selection of pre and post smoothing parameters} \label{sec:prepostsmoothing}

\begin{figure}
  \resizebox{\hsize}{!}{\includegraphics{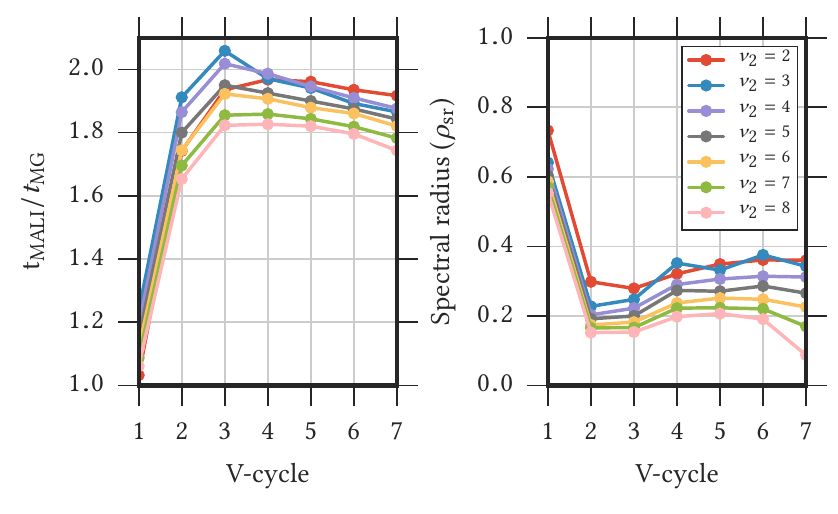}}
  \caption{Speed-up (left-hand panel) and spectral radius (right-hand panel) of multigrid as function of the number of V-cycles. The computations were performed using column $x=0,y=271$ from Model 1 as a one-dimensional atmosphere and the three-level \CaII\ model atom. The different colors indicate a different number of post-smoothing iterations $\nu_2$. The other multigrid parameters are: three grids, $\nu_1 = 2$, $\nu_3 = 32$, linear interpolation and full-weighting restriction.}
  \label{fig:smoothing_number_caIIK}
\end{figure}

The convergence and stability of the multigrid method depends on the selection of the multigrid parameters $\nu_1$,$\nu_2$ and $\nu_3$ (see Sec.~\ref{sec:multigrid}). To ensure a good initial smoothing before the first V-cycle, we define an additional parameter, $\nu_0$, which is the number of MALI iterations performed after initialization. This means that at the finest grid level we perform $\nu_0$ iterations during the first V-cycle, but use $\nu_1$ during all later V-cycles. 

In order to analyze the multigrid behavior as function of the $\nu$ parameters we use two metrics. The first is the spectral radius. The other metric is the speed-up $S_i$, which we define as the ratio
\begin{equation}
S_i = t_\mathrm{MALI}/t^i_\mathrm{MG},
\end{equation}
where $t^i_\mathrm{MG}$ is the computing time required to perform $i$ V-cycles (with an associated error $E_{i,\infty}$), and $t_\mathrm{MALI}$ the computing time to reach the same error. A small spectral radius means one needs fewer V-cycles to reach convergence, but this does not necessarily mean that $S_i$ also increases because one might need many iterations (as defined by the value of $\nu_1$,$\nu_2$ and $\nu_3$) within a V-cycle to obtain the small spectral radius.

We found that every atmosphere and model atom behaves differently, and it is impossible to give a set of parameters that always provides the fastest stable multigrid iteration. Empirically we found that two to four pre-smoothings ($\nu_1=2$\,--\,$4$) worked well in all our test cases, but for Model 1 and 2 with the hydrogen atom we required extra initial smoothing for the first cycle ($\nu_0 \geq 15$). At the coarsest grid, we found that $\nu_3 \approx 32$ is a good number, independent of the problem. Iterating further did not significantly influence the convergence speed or stability. 

The most sensitive parameter is the post-smoothing number. Our tests with the \CaII\ atoms in Model 1 and FAL-C showed stable and fast convergence with $\nu_2 = 2$, but for Model 2 we used $\nu_2 = 8$ to avoid excessive occurrence of negative populations. In Fig.~\ref{fig:smoothing_number_caIIK} we show the speedup and spectral radius in a 1D calculation where we used a column from Model 1 as a plane-parallel 1D atmosphere. The spectral radius decreases with increasing $\nu_2$, but the speedup gets smaller owing to the increase in computational cost per V-cycle.

For hydrogen we found that one needs many more post smoothing iterations ($\nu_2=25$). If a smaller number is chosen then the iteration at the coarsest grid tends to produce large areas with negative populations. We found that this is related to the smoothness of the error at the finest grid. In Figure~ \ref{fig:negative_pop_coarse_grid} we illustrate this. The two left-hand columns show the error for the first two levels of our hydrogen atom at the start of the third V-cycle for computations with a different number of post-smoothings ($\nu_2=2$, $10$ and $20$). The error gets smoother with increasing $\nu_2$. The right-hand columns show the sign of the level populations and the maximum correction as function of the iteration number at the coarsest grid during the third V-cycle. For $\nu_2=2$ a negative population occurs already during the first coarse-grid iteration and this remains so for all other iterations. For $\nu_2=10$ negative populations occur between iteration 3 and 17 after which all populations become positive. For $\nu_2=20$ negative populations occur only during iteration 4 to 12. For comparison we show in the fourth row iterations at the coarsest grid for $\nu_2=2$ where we solve Eq.~\ref{eq:fundemental_eq} instead of Eq.~\ref{coarse_eq}, i.e., we perform standard MALI iterations instead of coarse-grid iterations. Here no negative populations occur, demonstrating that negative populations occur owing to the right-hand-side of Eq.~\ref{coarse_eq} and not owing to the coarse grid spacing.

\subsection{Spectral radius and convergence criteria} \label{sec:spandconv}

The spectral radius for the three-level \CaII\ with MALI one-grid iteration in Model 1 is $\rho_\mathrm{sr}=0.971$, for and \HI\ atoms it is $\rho_\mathrm{sr}= 0.9923$. An optimal multigrid method should achieve a spectral radius of 0.1-0.2 \citepads{multigrid_trottenberk}. We obtained a spectral radius in the range of 0.3-0.4 with the pre and post-smoothing parameters as discussed in Sec~\ref{sec:prepostsmoothing} using Eq.~\ref{eq:spectral_radius_def}. 

The spectral radius of one-grid MALI iterations increases with decreasing grid spacing
\citepads{1986JQSRT..35..431O}.
Combined with Eq. \ref{eq:error_approx}, this means that one needs to iterate longer and longer to reach the same level of accuracy as the grid spacing gets finer. In contrast, the spectral radius of multigrid remains constant irrespective of the grid spacing
\citepads[][]{1991A&A...242..290S,1997A&A...324..161F,multigrid_trottenberk},
so that the speed-up of multigrid is expected to increase with decreasing grid spacing.

\begin{figure*}
  \begin{minipage}{\textwidth}
      \includegraphics[width=\textwidth]{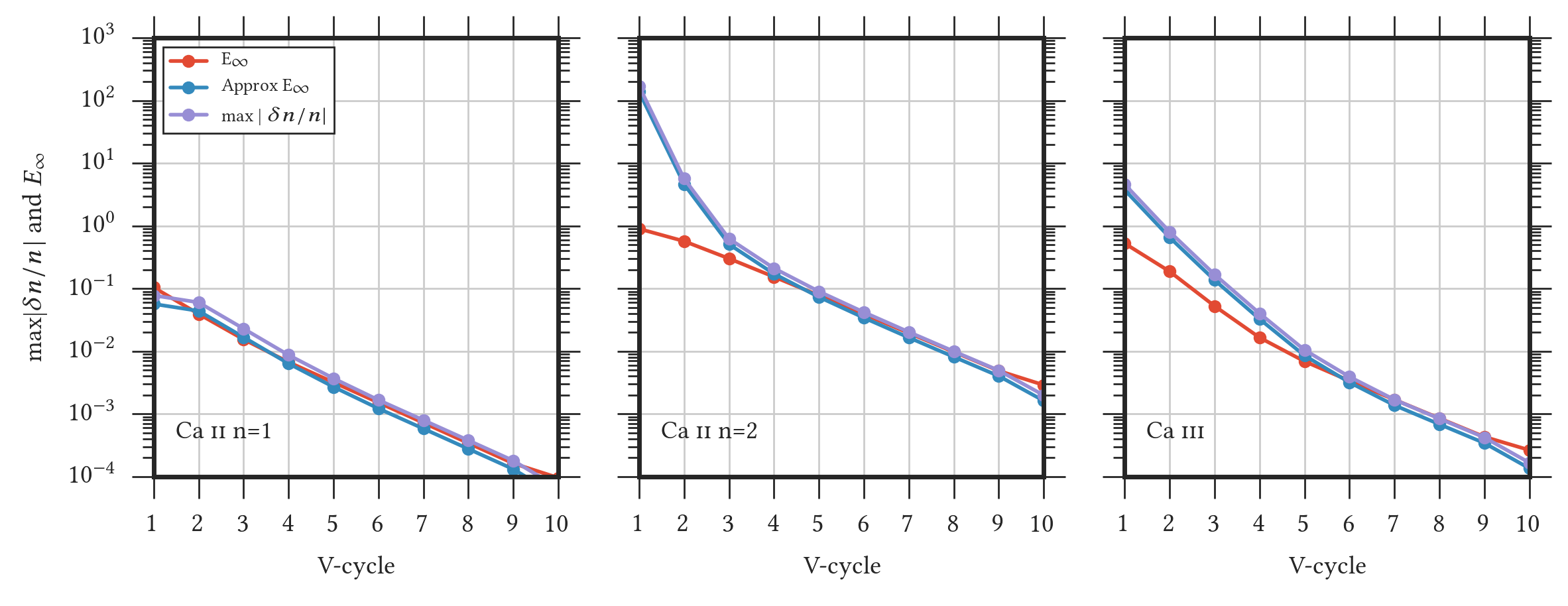}\hfil
     \includegraphics[width=\textwidth]{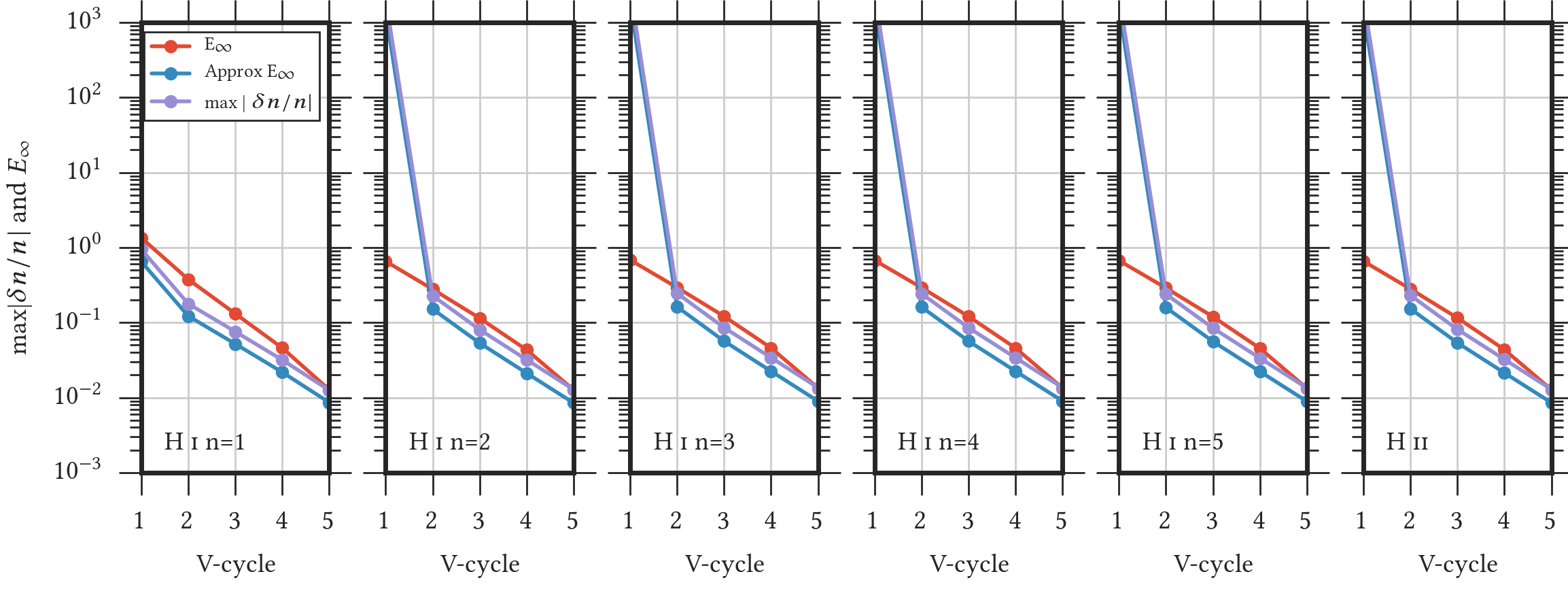}\hfil
  \end{minipage}
  \caption{
  Convergence behavior of multigrid showing the maximum relative error and the maximum relative error estimated by Eq.~\ref{eq:error_approx} for each V-cycle, as well as the maximum relative change in population per V-cycle. The upper panels shows the three-level \CaII\ atom and lower panels shows \HI, both for Atmosphere model 1. For three-level \CaII\ we used  $\nu_1 = 2$, $\nu_2 = 2$, $\nu_3=32$ and for \HI\ we used  $\nu_0=15$, $\nu_1 = 2$, $\nu_2 = 25$, $\nu_3=32$.}
  \label{fig:error_vs_relativecor_vcycle}
\end{figure*}

The relation between the maximum relative correction and the error (Eq.~\ref{eq:error_approx}) is valid for one-grid MALI iteration
\citepads{1994A&A...292..599A}.
\citetads{1997A&A...324..161F}
showed that it holds for multigrid in a smooth 2D atmosphere too. The 3D multigrid computations presented in 
\citetads{2013A&A...557A.143S} 
did not test the validity of the relation in 3D. Therefore we tested the validity in atmosphere Model 1. The results are shown in Figure~\ref{fig:error_vs_relativecor_vcycle}, where we compare the true error, the error estimate from Eq.~\ref{eq:error_approx} and the maximum relative correction.
For \CaII\ we find that the formula still holds. We also find that the maximum relative correction is equal to the error. This is accidental, because the spectral radius is close to 0.5. For different pre-and post-smoothing parameters a different result would be obtained. Also note that the maximum relative correction in Eq.~\ref{eq:error_approx} must be for the whole V-cycle, and not for the individual MALI iterations in the post-smoothing at the finest grid.
In the hydrogen case the approximation underestimates the true error by a factor 2. Again the maximum relative correction is just similar to the error because of the spectral radius being close to 0.5.
The \CaII\ computation reaches the linear regime after five V-cycles. For \HI\ it requires only one cycle. This is owing to the much higher number of post-smoothing iterations for hydrogen: $\nu_{2,\mathrm{H I}} = 25$ and $\nu_{2,\mathrm{Ca II}} = 2$.

\section{Results} \label{sec:results}

\begin{figure*}
  \begin{minipage}{\textwidth}
      \includegraphics[width=\textwidth]{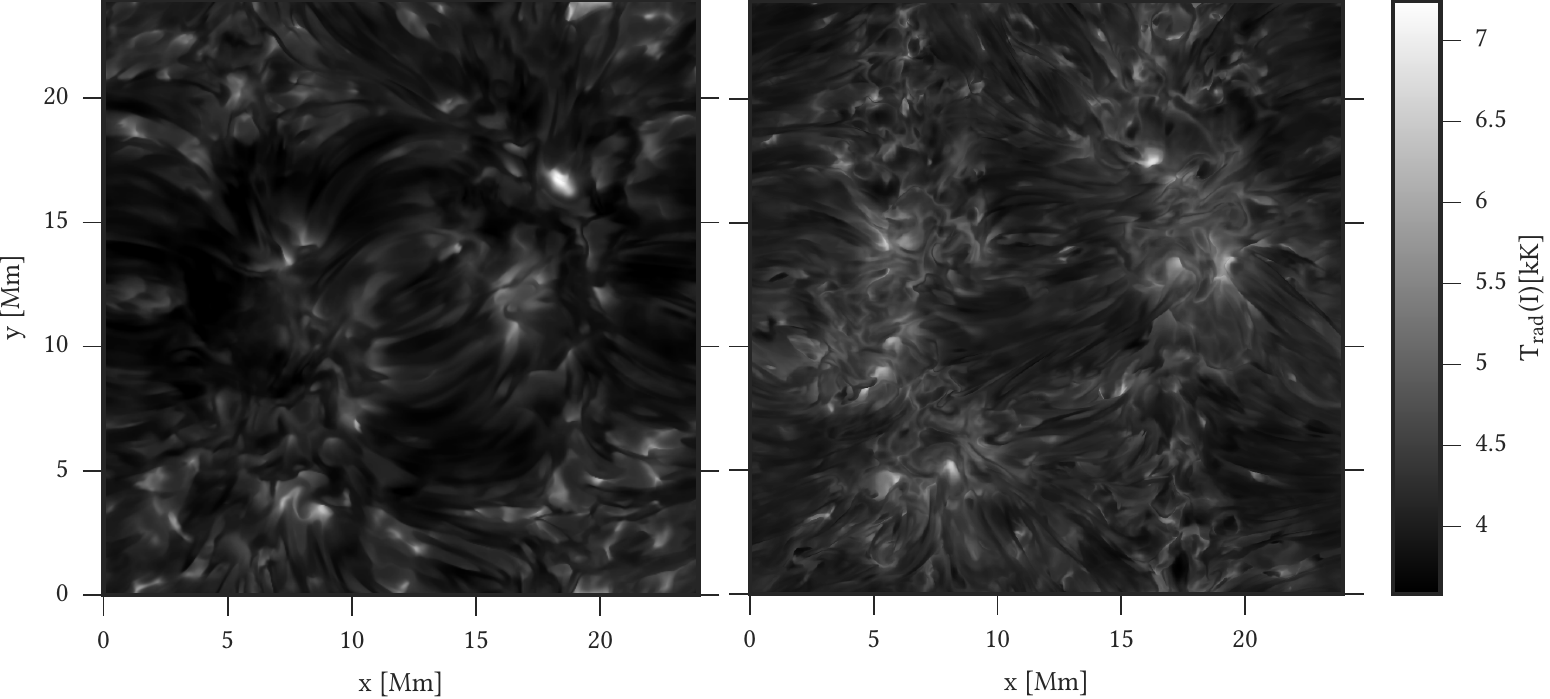}\hfil
  \end{minipage}
  \caption{The intensity of the \CaII\ K line core computed from atmosphere Model 1 (left) and Model 2 (right) at $\mu_z=1$. The intensity is shown as the brightness temperature $T_\mathrm{rad}$ computed from $B_{\nu}(T_{\mathrm{rad}}) = I_{\nu}$.  }
  \label{fig:trad_bifrost_atm}
\end{figure*}

\begin{table*}
  \caption{Summary of 3D non-LTE multigrid runs.}
  \label{tab:final_runs}
  \centering
    \begin{tabular}{l l l r l}
    \hline \hline
     Atmosphere & Atom & Method & $\nu_\mathrm{FMG}$,$\nu_0$,$\nu_1$,$\nu_2$,$\nu_3$ & Speed-up \\
    \hline
    Model 1 & 3-level \CaII\ & MALI              &                    &1 \\
                   &      & 3-grid MG      & --, 0, 2, 2, 32   & 3.3 \\		
                    &     & 3-grid full-MG& 16, 0, 2, 2, 32  & 6 \\		
    &6-level    \HI\ & MALI            &                      & 1 \\	
    &                     & 3-grid MG    & --, 15, 2, 25, 32 & 4.5\\	
    Model 2 & 6-level \CaII\	 & 3-grid MG & --, 20, 4, 8, 32 & unknown \\
       \hline
     \end{tabular}
 \end{table*}

\begin{figure*}[!t]
  \begin{minipage}{\textwidth}
    \includegraphics[width=0.5\textwidth]{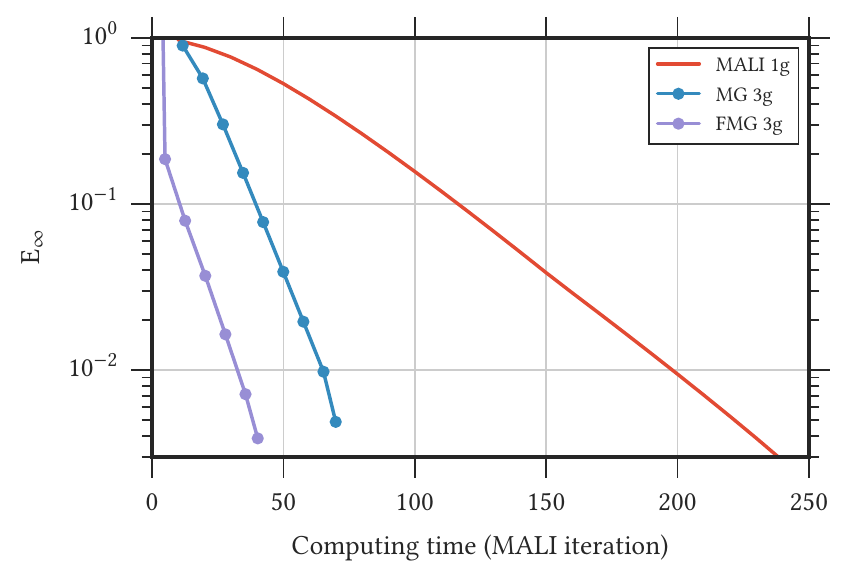}\hfil
    \includegraphics[width=0.5\textwidth]{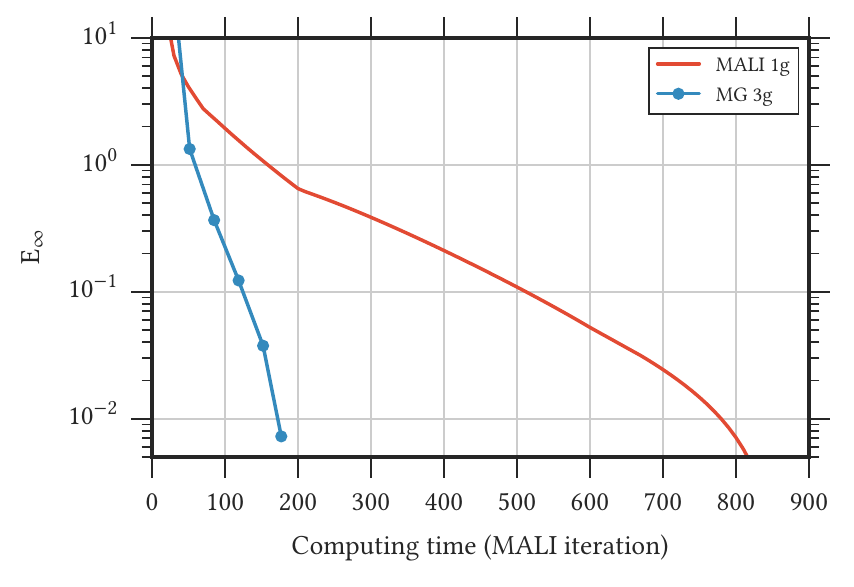}\hfil
  \end{minipage}
  \caption{
  Convergence behavior for MALI and multigrid for three-level \CaII\ (left-hand panel) and  hydrogen atom (right-hand panel) for atmosphere Model 1. The computation was performed with three grids, full-weighting restriction and trilinear interpolation. For three-level \CaII\ we used  $\nu_1 = 2$, $\nu_2 = 2$, $\nu_3=32$ and for \HI\ we used  $\nu_0=15$,$\nu_1 = 2$, $\nu_2 = 25$, $\nu_3=32$.}
  \label{fig:convergence_error_504_504_497}
\end{figure*}

\begin{figure}
    \centering
      \resizebox{\hsize}{!}{\includegraphics[]{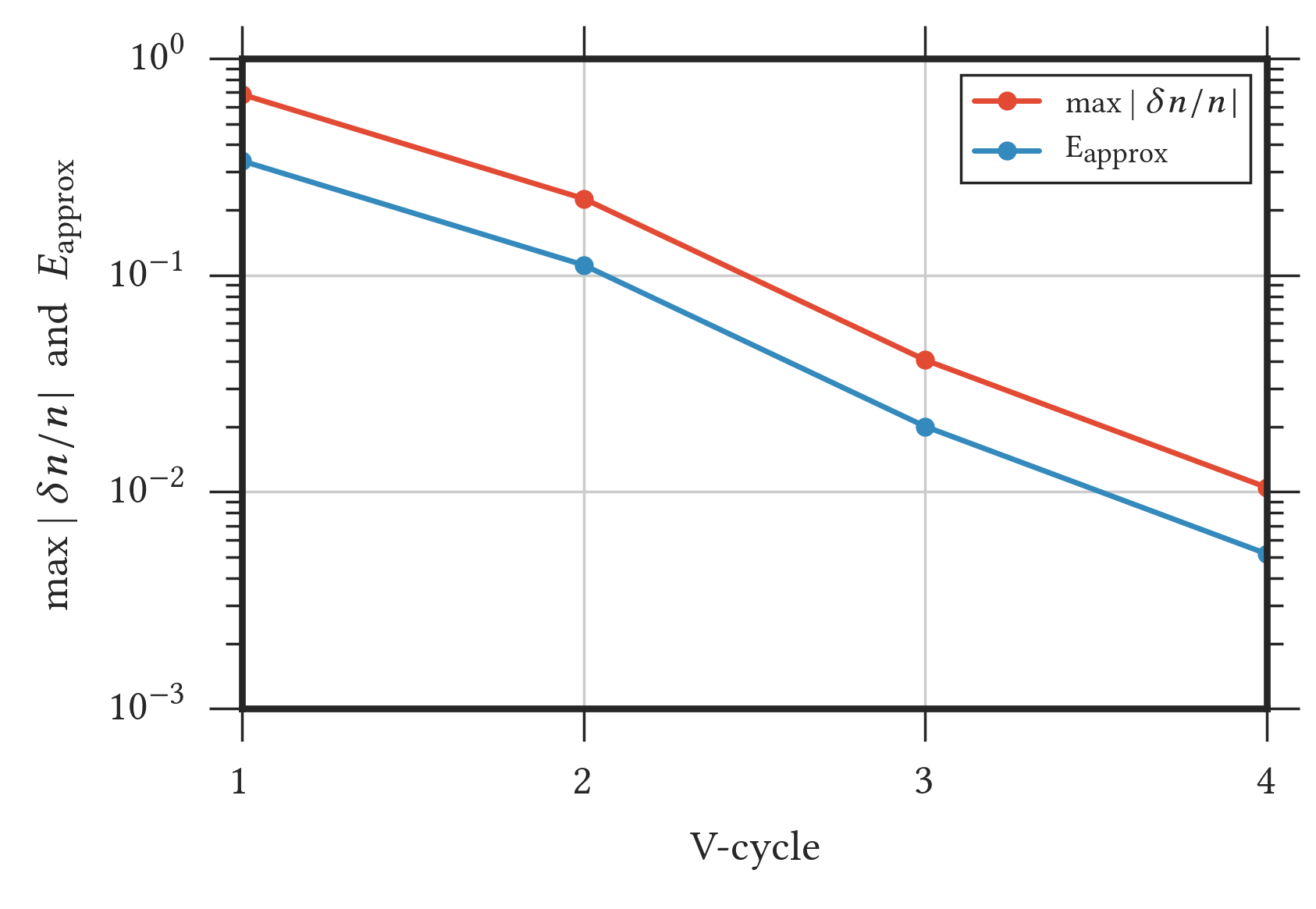}}
    \vspace{1ex}
  \caption{ The convergence properties for atmosphere Model 2 with six-level \CaII\ as function of V-cycle. The $E_{\mathrm{approx}}$ is calculated based on the relation between the spectral radius and maximum relative population change (Eq. \ref{eq:error_approx}). The computation was performed with three grids, full-weighting, linear interpolation using $\nu_0=20$, $\nu_1=4$, $\nu_2=8$, $\nu_3=32$.}
\end{figure}

In Table~\ref{tab:final_runs} we show a summary of our 3D non-LTE multigrid computations. We obtained a speed-up of $3.3$  for \CaII\ and $4.5$ for \HI\ with atmosphere Model 1 and standard multigrid. Using full multigrid we obtain a speed-up of $6$ for \CaII, an improvement of a factor $1.8$ compared to ordinary multigrid. A similar result for a smooth atmosphere was found by 
\citetads{1997A&A...324..161F}.
In the hydrogen case we needed 25 post-smoothing iterations to avoid too many occurrences of negative populations. This considerably reduced the speed-up. The convergence behavior of these three runs is shown in Figure \ref{fig:convergence_error_504_504_497}.

We also performed multigrid computation in atmosphere Model 2, to test whether multigrid can handle a finer grid spacing ($\Delta_{x,y}=32$~km, $\Delta_z=13$~km), and a different atmosphere. Obtaining a reference  solution for this model was too computationally expensive. Therefore, we estimate the convergence based on Eq. \ref{eq:error_approx}, which we showed to be accurate for \CaII\ in Model 1 (see Fig.~\ref{fig:error_vs_relativecor_vcycle}). The computation with Model 2 proves that 3D non-LTE radiative transfer using multigrid can handle different atmospheres and grid spacings. 

Figure \ref{fig:trad_bifrost_atm} shows the brightness temperature in the \CaII\ K line core for Model 1 and 2. The image computed from Model 2 shows substantially more fine-structure. This highlights the need for higher-resolution radiation-MHD simulations and the capacity to efficiently compute synthetic diagnostics using 3D non-LTE radiative transfer from them.

\section{Summary and conclusions} \label{sec:conclusions}

In this article, we investigated whether a non-linear multigrid method can be applied to 3D non-LTE radiative transfer problems using snapshots from radiation-MHD simulations as input. We showed that this is indeed the case: we obtain a speedup between 4.5 and 6 for a $504\times504\times496$ atmosphere (48~km horizontal and 19~km vertical resolution) computed with the radiation-MHD code Bifrost for \CaII\ and \HI\ atoms. 
We also solve the 3D non-LTE radiative transfer problem using multigrid on for a  $768\times768\times768$ (32~km horizontal and 13~km vertical resolution) Bifrost snapshot.
We conclude that the multigrid method can handle model atmospheres with very strong gradients in atmospheric parameters as well as strongly scattering lines such as hydrogen Ly$\alpha$ and \CaII~H\&K.
 
The implementation of multigrid requires choosing suitable numerical methods for the sub-tasks. We investigated several of those choices using 1D and 3D test computations. In summary our findings are:
\begin{itemize}
\item use full-weighting as restriction operator
\item linear interpolation works well. Third-order Hermite interpolations appears to give higher convergence in 1D problems and should be investigated in 3D problems.
\item one should interpolate the relative correction to the populations, and not the absolute correction.
\item use full-multigrid if possible. 
\item initializing with the zero-radiation approximation is substantially better than initializing with LTE.
\item three-grid iteration converges faster than four-grid iteration. 
\item the coarse-grid iterations can converge to a negative solution. This is not a problem for isolated grid points, but if extended regions have negative populations one should increase the number of post-smoothing iterations.
\item each atom and atmosphere requires some testing to find the optimal number of pre-smoothings and post-smoothings. Our findings in Table~\ref{tab:final_runs} can be used as a starting point. 
\end{itemize}

Since each problem is unique, other atmospheres and atoms could require different approaches. Therefore, the multigrid method should be implemented into radiative transfer codes in a modular way so that methods can be easily changed. 

We did not obtain the high convergence rate (as measured in spectral radius of the multigrid iteration) as reported in 
\citetads{1997A&A...324..161F}. 
This has two reasons: First, these authors used a static smooth 2D with a weak horizontal temperature inhomogeneity and no vertical temperature gradient, while we are using moving atmospheres with very large gradients in all atmospheric parameters. Second, they use Gauss-Seidel (GS) iterations, while we use Jacobi iteration in Multi3D. The smoothing properties and the convergence speed of GS iterations are superior to Jacobi iteration. Unfortunately no MPI-parallelization scheme for GS iteration that scales well to thousands of computing cores exist, and we are forced to use Jacobi iteration. The lower convergence speed per iteration for Jacobi iteration can fortunately be offset by increasing the number of computing cores, but ideally one should develop an efficient parallel GS iteration scheme. A similar conclusion was reached by
\citetads{2013A&A...557A.143S}.
So far we have tested our multigrid method only using complete frequency redistribution. Because partial frequency redistribution (PRD) can increase the computing time in non-LTE problems by more than an order of magnitude, the obvious next step will be to test the combination of 3D PRD 
\citepads{2016arXiv160605180S}
and multigrid in realistic use-cases.
{In this paper we have used small model atoms with up to six levels. Multigrid method should also be tested with more complex atoms, such as \ion{Fe}{I}, which are important for stellar photospheric abundance determinations \citepads[e.g.][]{2016A&A...597A...6N}.}

The speed-up factor of multigrid increases with decreasing grid spacing. With the ever-increasing grid sizes and decreasing grid spacing of solar and stellar radiation-MHD models will make the use multigrid methods in 3D non-LTE radiative transfer more and more desirable to keep computing costs manageable.

\begin{acknowledgements}
  Computations were performed on resources provided by the Swedish National
  Infrastructure for Computing (SNIC) at the National Supercomputer Centre at
  Link\"oping University, at the High Performance Computing Center North at
  Ume\aa\ University, and at the PDC Centre for High Performance Computing (PDC-HPC)
  at the Royal Institute of Technology in Stockholm. JPB and JL thank Mats Carlsson and Viggo Hansteen for providing a $768^3$ Bifrost snapshot. JPB thanks Andrii Sukhorukov for answering questions regarding radiative transfer.
\end{acknowledgements}

\bibliographystyle{aa} 

\begin{thebibliography}{47}
\expandafter\ifx\csname natexlab\endcsname\relax\def\natexlab#1{#1}\fi

\bibitem[{{Amarsi} {et~al.}(2016){Amarsi}, {Asplund}, {Collet}, \&
  {Leenaarts}}]{2016MNRAS.455.3735A}
{Amarsi}, A.~M., {Asplund}, M., {Collet}, R., \& {Leenaarts}, J. 2016, \mnras,
  455, 3735

\bibitem[{{Asplund} {et~al.}(2003){Asplund}, {Carlsson}, \&
  {Botnen}}]{2003A&A...399L..31A}
{Asplund}, M., {Carlsson}, M., \& {Botnen}, A.~V. 2003, \aap, 399, L31

\bibitem[{{Asplund} {et~al.}(2004){Asplund}, {Grevesse}, {Sauval}, {Allende
  Prieto}, \& {Kiselman}}]{2004A&A...417..751A}
{Asplund}, M., {Grevesse}, N., {Sauval}, A.~J., {Allende Prieto}, C., \&
  {Kiselman}, D. 2004, \aap, 417, 751

\bibitem[{{Auer}(2003)}]{2003ASPC..288....3A}
{Auer}, L. 2003, in Astronomical Society of the Pacific Conference Series, Vol.
  288, Stellar Atmosphere Modeling, ed. I.~{Hubeny}, D.~{Mihalas}, \&
  K.~{Werner}, 3

\bibitem[{{Auer} {et~al.}(1994){Auer}, {Bendicho}, \& {Trujillo
  Bueno}}]{1994A&A...292..599A}
{Auer}, L., {Bendicho}, P.~F., \& {Trujillo Bueno}, J. 1994, \aap, 292, 599

\bibitem[{Brandt(1977)}]{brandt1977multi}
Brandt, A. 1977, Mathematics of computation, 31, 333

\bibitem[{{Cannon}(1973)}]{1973ApJ...185..621C}
{Cannon}, C.~J. 1973, \apj, 185, 621

\bibitem[{Carlson {et~al.}(1963)Carlson, Adler, Fernbach, \&
  Rotenberg}]{carlson1963methods}
Carlson, B., Adler, B., Fernbach, S., \& Rotenberg, M. 1963, B. Alder, S.
  Fernbach, \& M. Rotenberg (New York: Academic), 1

\bibitem[{{Carlsson} {et~al.}(2016){Carlsson}, {Hansteen}, {Gudiksen},
  {Leenaarts}, \& {De Pontieu}}]{2016A&A...585A...4C}
{Carlsson}, M., {Hansteen}, V.~H., {Gudiksen}, B.~V., {Leenaarts}, J., \& {De
  Pontieu}, B. 2016, \aap, 585, A4

\bibitem[{{Carlsson} \& {Leenaarts}(2012)}]{2012A&A...539A..39C}
{Carlsson}, M. \& {Leenaarts}, J. 2012, \aap, 539, A39

\bibitem[{{de la Cruz Rodr{\'{\i}}guez} {et~al.}(2013){de la Cruz
  Rodr{\'{\i}}guez}, {De Pontieu}, {Carlsson}, \& {Rouppe van der
  Voort}}]{2013ApJ...764L..11D}
{de la Cruz Rodr{\'{\i}}guez}, J., {De Pontieu}, B., {Carlsson}, M., \& {Rouppe
  van der Voort}, L.~H.~M. 2013, \apjl, 764, L11

\bibitem[{{Fabiani Bendicho} {et~al.}(1997){Fabiani Bendicho}, {Trujillo
  Bueno}, \& {Auer}}]{1997A&A...324..161F}
{Fabiani Bendicho}, P., {Trujillo Bueno}, J., \& {Auer}, L. 1997, \aap, 324,
  161

\bibitem[{{Fontenla} {et~al.}(1993){Fontenla}, {Avrett}, \&
  {Loeser}}]{1993ApJ...406..319F}
{Fontenla}, J.~M., {Avrett}, E.~H., \& {Loeser}, R. 1993, \apj, 406, 319

\bibitem[{{Gudiksen} {et~al.}(2011){Gudiksen}, {Carlsson}, {Hansteen}, {Hayek},
  {Leenaarts}, \& {Mart{\'{\i}}nez-Sykora}}]{2011A&A...531A.154G}
{Gudiksen}, B.~V., {Carlsson}, M., {Hansteen}, V.~H., {et~al.} 2011, \aap, 531,
  A154

\bibitem[{Hackbush(1985)}]{hackbush}
Hackbush, W. 1985, Multi-Grid Methods and Applications (Springer Berlin
  Heidelberg)

\bibitem[{{Holzreuter} \& {Solanki}(2013)}]{2013A&A...558A..20H}
{Holzreuter}, R. \& {Solanki}, S.~K. 2013, \aap, 558, A20

\bibitem[{{Holzreuter} \& {Solanki}(2015)}]{2015A&A...582A.101H}
{Holzreuter}, R. \& {Solanki}, S.~K. 2015, \aap, 582, A101

\bibitem[{{Ibgui} {et~al.}(2013){Ibgui}, {Hubeny}, {Lanz}, \&
  {Stehl{\'e}}}]{2013A&A...549A.126I}
{Ibgui}, L., {Hubeny}, I., {Lanz}, T., \& {Stehl{\'e}}, C. 2013, \aap, 549,
  A126

\bibitem[{{Judge} {et~al.}(2015){Judge}, {Kleint}, {Uitenbroek}, {Rempel},
  {Suematsu}, \& {Tsuneta}}]{2015SoPh..290..979J}
{Judge}, P.~G., {Kleint}, L., {Uitenbroek}, H., {et~al.} 2015, \solphys, 290,
  979

\bibitem[{{Kunasz} \& {Auer}(1988)}]{1988JQSRT..39...67K}
{Kunasz}, P. \& {Auer}, L.~H. 1988, \jqsrt, 39, 67

\bibitem[{{Leenaarts} \& {Carlsson}(2009)}]{2009ASPC..415...87L}
{Leenaarts}, J. \& {Carlsson}, M. 2009, in Astronomical Society of the Pacific
  Conference Series, Vol. 415, The Second Hinode Science Meeting: Beyond
  Discovery-Toward Understanding, ed. B.~{Lites}, M.~{Cheung}, T.~{Magara},
  J.~{Mariska}, \& K.~{Reeves}, 87

\bibitem[{{Leenaarts} {et~al.}(2009){Leenaarts}, {Carlsson}, {Hansteen}, \&
  {Rouppe van der Voort}}]{2009ApJ...694L.128L}
{Leenaarts}, J., {Carlsson}, M., {Hansteen}, V., \& {Rouppe van der Voort}, L.
  2009, \apjl, 694, L128

\bibitem[{{Leenaarts} {et~al.}(2012){Leenaarts}, {Carlsson}, \& {Rouppe van der
  Voort}}]{2012ApJ...749..136L}
{Leenaarts}, J., {Carlsson}, M., \& {Rouppe van der Voort}, L. 2012, \apj, 749,
  136

\bibitem[{{Leenaarts} {et~al.}(2013{\natexlab{a}}){Leenaarts}, {Pereira},
  {Carlsson}, {Uitenbroek}, \& {De Pontieu}}]{2013ApJ...772...89L}
{Leenaarts}, J., {Pereira}, T.~M.~D., {Carlsson}, M., {Uitenbroek}, H., \& {De
  Pontieu}, B. 2013{\natexlab{a}}, \apj, 772, 89

\bibitem[{{Leenaarts} {et~al.}(2013{\natexlab{b}}){Leenaarts}, {Pereira},
  {Carlsson}, {Uitenbroek}, \& {De Pontieu}}]{2013ApJ...772...90L}
{Leenaarts}, J., {Pereira}, T.~M.~D., {Carlsson}, M., {Uitenbroek}, H., \& {De
  Pontieu}, B. 2013{\natexlab{b}}, \apj, 772, 90

\bibitem[{{L{\'e}ger} {et~al.}(2007){L{\'e}ger}, {Chevallier}, \&
  {Paletou}}]{2007A&A...470....1L}
{L{\'e}ger}, L., {Chevallier}, L., \& {Paletou}, F. 2007, \aap, 470, 1

\bibitem[{{Loukitcheva} {et~al.}(2015){Loukitcheva}, {Solanki}, {Carlsson}, \&
  {White}}]{2015A&A...575A..15L}
{Loukitcheva}, M., {Solanki}, S.~K., {Carlsson}, M., \& {White}, S.~M. 2015,
  \aap, 575, A15

\bibitem[{{Nordlander} {et~al.}(2016){Nordlander}, {Amarsi}, {Lind}, {Asplund},
  {Barklem}, {Casey}, {Collet}, \& {Leenaarts}}]{2016A&A...597A...6N}
{Nordlander}, T., {Amarsi}, A.~M., {Lind}, K., {et~al.} 2016, \aap, 597, A6

\bibitem[{{Olson} {et~al.}(1986){Olson}, {Auer}, \&
  {Buchler}}]{1986JQSRT..35..431O}
{Olson}, G.~L., {Auer}, L.~H., \& {Buchler}, J.~R. 1986, \jqsrt, 35, 431

\bibitem[{{Paletou} \& {Anterrieu}(2009)}]{2009A&A...507.1815P}
{Paletou}, F. \& {Anterrieu}, E. 2009, \aap, 507, 1815

\bibitem[{{Pereira} {et~al.}(2009){Pereira}, {Asplund}, \&
  {Kiselman}}]{2009A&A...508.1403P}
{Pereira}, T.~M.~D., {Asplund}, M., \& {Kiselman}, D. 2009, \aap, 508, 1403

\bibitem[{{Pereira} {et~al.}(2015){Pereira}, {Carlsson}, {De Pontieu}, \&
  {Hansteen}}]{2015ApJ...806...14P}
{Pereira}, T.~M.~D., {Carlsson}, M., {De Pontieu}, B., \& {Hansteen}, V. 2015,
  \apj, 806, 14

\bibitem[{{Rathore} \& {Carlsson}(2015)}]{2015ApJ...811...80R}
{Rathore}, B. \& {Carlsson}, M. 2015, \apj, 811, 80

\bibitem[{{Rathore} {et~al.}(2015){Rathore}, {Carlsson}, {Leenaarts}, \& {De
  Pontieu}}]{2015ApJ...811...81R}
{Rathore}, B., {Carlsson}, M., {Leenaarts}, J., \& {De Pontieu}, B. 2015, \apj,
  811, 81

\bibitem[{{Rybicki} \& {Hummer}(1991)}]{1991A&A...245..171R}
{Rybicki}, G.~B. \& {Hummer}, D.~G. 1991, \aap, 245, 171

\bibitem[{{Rybicki} \& {Hummer}(1992)}]{1992A&A...262..209R}
{Rybicki}, G.~B. \& {Hummer}, D.~G. 1992, \aap, 262, 209

\bibitem[{{Steiner}(1991)}]{1991A&A...242..290S}
{Steiner}, O. 1991, \aap, 242, 290

\bibitem[{{Sukhorukov} \& {Leenaarts}(2016)}]{2016arXiv160605180S}
{Sukhorukov}, A.~V. \& {Leenaarts}, J. 2016, ArXiv e-prints
  [\eprint[arXiv]{1606.05180}]

\bibitem[{Trottenberg {et~al.}(2000)Trottenberg, Oosterlee, \&
  Schuller}]{multigrid_trottenberk}
Trottenberg, U., Oosterlee, C.~W., \& Schuller, A. 2000, Multigrid (Academic
  press)

\bibitem[{{Trujillo Bueno} \& {Fabiani Bendicho}(1995)}]{1995ApJ...455..646T}
{Trujillo Bueno}, J. \& {Fabiani Bendicho}, P. 1995, \apj, 455, 646

\bibitem[{{Trujillo Bueno} \& {Shchukina}(2007)}]{2007ApJ...664L.135T}
{Trujillo Bueno}, J. \& {Shchukina}, N. 2007, \apjl, 664, L135

\bibitem[{Tsitsiklis {et~al.}(1988)}]{tsitsiklis1988comparison}
Tsitsiklis, J.~N. {et~al.} 1988, A comparison of Jacobi and Gauss-Seidel
  parallel iterations (Massachusetts Institute of Technology, Laboratory for
  Information and Decision Systems)

\bibitem[{{Uitenbroek}(2001)}]{2001ApJ...557..389U}
{Uitenbroek}, H. 2001, \apj, 557, 389

\bibitem[{{Uitenbroek}(2006)}]{2006ApJ...639..516U}
{Uitenbroek}, H. 2006, \apj, 639, 516

\bibitem[{{{\v S}t{\v e}p{\'a}n} \& {Trujillo
  Bueno}(2013)}]{2013A&A...557A.143S}
{{\v S}t{\v e}p{\'a}n}, J. \& {Trujillo Bueno}, J. 2013, \aap, 557, A143

\bibitem[{{{\v S}t{\v e}p{\'a}n} {et~al.}(2015){{\v S}t{\v e}p{\'a}n},
  {Trujillo Bueno}, {Leenaarts}, \& {Carlsson}}]{2015ApJ...803...65S}
{{\v S}t{\v e}p{\'a}n}, J., {Trujillo Bueno}, J., {Leenaarts}, J., \&
  {Carlsson}, M. 2015, \apj, 803, 65

\bibitem[{{Vath}(1994)}]{1994A&A...284..319V}
{Vath}, H.~M. 1994, \aap, 284, 319

\end{thebibliography}

\end{document}